\documentclass[twocolumn,showpacs,superscriptaddress,prb,notitlepage]{revtex4-1}
\usepackage{graphicx}
\usepackage{color}
\usepackage{amsmath,amsthm,amssymb}
\usepackage{braket}
\usepackage[colorlinks,citecolor=blue,linkcolor=red]{hyperref}
\usepackage{multirow}
\usepackage{ulem}
\usepackage{cancel}

\newcommand{\red}{\color{red}}

\begin{document}
 
\title{Gor'kov-Hedin-Baym Equations for Quantum Many-Body Systems with Spin-Dependent Interactions}

\author{Christopher Lane}
\email{laneca@lanl.gov}
\affiliation{Theoretical Division, Los Alamos National Laboratory, Los Alamos, New Mexico 87545, USA}

\date{\today} 
\begin{abstract}
Driven by the need to understand and determine the presence of non-trivial superconductivity in real candidate materials, we present a generalized set of self-consistent Gor'kov-Hedin-Baym equations with spin dependent electron-electron and electron-phonon interactions. This extends Hedin's original equations to treat quantum many-body systems where electronic and lattice correlations along with relativistic effects coexist on the same footing with superconductivity. The leading order self-energies yields a generalization of the Migdal-Eliashberg theory and by iterating this set of equations generalized ladder vertex corrections naturally emerge.
\end{abstract}


\maketitle 

\section{Introduction}\label{sec:intro}
An increasing number of emerging technologies rely on exotic forms of superconductivity that arise at the intersection of correlations and relativistic effects. Fault tolerant quantum computing\cite{sato2017topological,nayak2008non}, spintronics\cite{eschrig2015spin,eschrig2011spin}, sensing\cite{alidoust2013singlet}, and quantum switching\cite{giazotto2010superconducting} are important examples not only due to their technological promise but also owing to the fundamental questions they raise. Crucial to understanding and designing the properties of these systems we must be able to theoretically describe the interplay of lattice, charge, spin, and orbital degrees of freedom and their mutual coupling in the presence of superconductivity. Despite scientific and technological needs, progress in developing techniques for describing superconductivity beyond the mean-field theory in the presences of strong relativistic effects has been slow. In particular, the Gor'kov Green's function treatment still awaits extension to strongly spin-orbit coupled systems. 

The landmark work of Bardeen, Cooper, and Schrieffer (BCS) beautifully elucidated the observations of H. Kamerlingh Onnes nearly 50 years earlier via the phonon mediated condensation of electron pairs\cite{cooper1956bound,bardeen1957microscopic,bardeen1957theory,allen1983theory}. Migdal and Eliashberg generalized BCS theory making it amendable for first-principle calculations\cite{migdal1958interaction,eliashberg1960interactions,eliashberg1961temperature} and by using the firm footing of quantum field theory\cite{zhu2016bogoliubov,gor2010developing,gor1958energy,gor1959microscopic,migdal1958interaction,eliashberg1960interactions,eliashberg1961temperature}, Gor'kov's Green's function based theory captures a host of additional physical effects, including spatially inhomogeneous problems, e.g., alloys, and the formation of magnetic field vortices with an applied external magnetic field\cite{abrikosov1959theory}. Most importantly it facilitates extensions that go beyond the weak coupling approximation of BCS and opens the door to new pairing mechanisms. 

Since the introduction of Gor'kov's Green's function, many material families have been discovered that appear to go beyond the weak coupling limit and incorporate a wide breath of fluctuations\cite{scalapino1986d,monthoux1992spin,bickers1987cdw} and collective modes\cite{kresin1988layer,hepting2018three,singh2022acoustic,allender1973model,jarrell1988charge,weber1988cu} present in complex correlated solids. To describe Cooper pair formation in these systems approximations such as Kohn-Luttinger\cite{luttinger1966new,kohn1965new}, Fluctuation-Exchange (FLEX)\cite{bickers1989conservingI,esirgen1997fluctuation,bickers1989conserving}, T-matrix\cite{keller1999thermodynamics}, RPA theory\cite{romer2015pairing,scalapino1986d,lane2022identifying}, and self-consistent schemes\cite{kita2011self}, along with nonperturbative treatments\cite{jiang2021ground,zheng2017stripe,lin1988pairing,mai2021orbital,oliveira1988density,linscheid2015hedin} have been introduced to help shed light on the gap symmetry in several material families\cite{scalapino1986d,scalapino2012common,kreisel2022superconducting,romer2019knight,hirschfeld2011gap,lane2022identifying,nandkishore2012chiral}. Despite this, the microscopic mechanism of unconventional superconductors is still under debate and there are persistent problems in describing the interplay of electronic, magnetic, and lattice vibrations on the same footing, which may give rise to feedback loops\cite{he2018rapid} and bootstrapping of two or more fluctuation channels\cite{fradkin2015colloquium}.  

The last 17 years have witnessed the rapid expansion of non-trivial physics that emerge from strong spin-orbit coupling. Typically, relativistic corrections such as spin-orbit coupling are considered a small perturbation on top the electronic states in a solid\cite{ashcroft1976solid}. However, as the atomic number of the constituent atomic species increases, relativistic corrections can become dominant, resulting in the striking qualitative effects present in topological quantum materials\cite{sobota2021angle,keimer2017physics}. When correlated electron physics is combined with strong spin-orbit coupling we gain access to a whole new space of exotic phases of matter that remain largely unexplored. Superconductivity arising from non-trivial topological quantum states has been of particular interest due to proposals of Cooper pairs with non-collinear spin textures\cite{tsutsumi2024topological} and special non-Abelian quasiparticles\cite{sato2017topological,qi2011topological} that enable new multifunctional devices and computing platforms. Since the theoretical approximations of the last 30 years were designed for scenarios where spin-orbit coupling is very weak or non-existent, it is quite challenging to examine the appearance of non-trivial superconductivity and its microscopic origin. 

Recently, spin-dependent interactions have shown to be crucial in describing the normal state of quantum many-body systems with strong spin-orbit coupling\cite{aryasetiawan2008,muller2019electron,nabok2021electron,scherpelz2016implementation}. Additionally, a study generalizing the RPA paramagnetic pairing interaction to include spin-orbit coupling describes the competition of trivial and non-trivial mixed parity states in monolayer transition metal dichalcogenides\cite{lane2022identifying}. However, such generalizations have yet to be incorporated into a fully self-consistent Gor'kov Green's function framework.

In this article, we present a generalized set of self-consistent Gor'kov-Hedin{\red-Baym} equations with spin dependent electron-electron and electron-phonon interactions that may arise from relativistic effects, such as spin-orbit coupling. This set of equations provides a firm basis for {\it ab initio} many-body perturbation theory calculations, where the leading order self-energies yields a generalization of the Migdal-Eliashberg theory and by iterating the Gor'kov-Hedin-Baym equations generalized ladder vertex corrections naturally emerge from the resulting spin-dependent vertex. 

\section{Gor'kov-Hedin-Baym equations with spin dependent Interactions}
The Hamiltonian for electrons in a vibrating lattice with spin dependent interactions and pairing fields is given by 
\begin{align}\label{eq:FullHamiltonian}
\hat{\mathcal{H}}&=\hat{\mathcal{T}}_{e}+\hat{\mathcal{T}}_{n}+\hat{\mathcal{U}}_{e-e}+\hat{\mathcal{U}}_{n-n}+\hat{\mathcal{U}}_{e-n}+\hat{\mathcal{P}}_{e},
\end{align}
where $\hat{\mathcal{T}}_{e}$ is the electronic kinetic energy
\begin{align}
\sum_{\alpha \beta } 
\int \mathrm{d}\mathbf{r} 
\hat{\psi}^{\dagger}_{\alpha }(\mathbf{r})
\left( -\frac{\hbar^2\nabla^2_{\mathbf{r}}}{2m_e}\delta_{\alpha\beta}+V^{soc}_{\alpha\beta}(\mathbf{r}) \right)
\hat{\psi}_{\beta }(\mathbf{r})
\end{align}
where $V^{soc}$ is the spin-orbit potential, and $m_e$ is the mass of the electron, $\hat{\mathcal{T}}_{n}$ is the nuclei kinetic energy, 
\begin{align}
\sum_{\kappa p}\frac{\hat{P}_{\kappa p}^2}{2M_\kappa}
\end{align}
with nuclear mass $M_\kappa$ of nucleus $\kappa$, $\hat{\mathcal{U}}_{e-e}$ is the spin-dependent electron-electron interaction 
\begin{align}
\frac{1}{2}\sum_{\substack{\alpha\beta\gamma\delta\\IJ}}\iint \mathrm{d}\mathbf{r}\mathrm{d}\mathbf{r}^{\prime} 
\hat{\psi}^{\dagger}_{\alpha }(\mathbf{r}) 
\hat{\psi}^{\dagger}_{\beta }(\mathbf{r}^\prime) 
\sigma^{I}_{\alpha\delta}v^{IJ}(\mathbf{r},\mathbf{r}^{\prime})\sigma^{J}_{\beta\gamma}
\hat{\psi}_{\gamma }(\mathbf{r}^\prime) 
\hat{\psi}_{\delta }(\mathbf{r}),
\end{align}
$\hat{\mathcal{U}}_{n-n}$ is the nuclei-nuclei interaction
\begin{align}
\frac{1}{2}\sum_{\substack{IJ}} 
\sum_{\kappa p \neq \kappa^\prime p^\prime}
X^I_{\kappa p}
v^{IJ}(\pmb{\tau}_{\kappa p},\pmb{\tau}_{\kappa^\prime p^\prime})
X^J_{\kappa^\prime p^\prime},
\end{align}
$\hat{\mathcal{U}}_{e-n}$ is the electron-nuclei interaction
\begin{align}
\sum_{\substack{I,J}}\iint \mathrm{d}\mathbf{r}\mathrm{d}\mathbf{r}^{\prime} 
n^{I}_{e}(\mathbf{r})
v^{IJ}(\mathbf{r},\mathbf{r}^{\prime})
n^{J}_{n}(\mathbf{r}^{\prime}),
\end{align}
and $\hat{\mathcal{P}}_{e}$ is the external pairing field
\begin{align}
\frac{1}{2}
\sum_{\alpha\beta}\iint &\mathrm{d}\mathbf{r}\mathrm{d}\mathbf{r}^{\prime} 
\hat{\psi}_{\alpha }(\mathbf{r})
\Delta_{\alpha\beta}(\mathbf{r},\mathbf{r}^\prime)  
\hat{\psi}_{\beta }(\mathbf{r}^\prime)\nonumber\\
&-
\frac{1}{2}
\sum_{\alpha\beta}\iint \mathrm{d}\mathbf{r}\mathrm{d}\mathbf{r}^{\prime} 
\hat{\psi}^{\dagger}_{\alpha }(\mathbf{r})
\bar{\Delta}_{\alpha\beta} (\mathbf{r},\mathbf{r}^\prime)
\hat{\psi}^{\dagger}_{\beta }(\mathbf{r}^\prime),
\end{align}
where the creation $\hat{\psi}^{\dagger}_{\alpha }(\mathbf{r}) $ and annihilation $\hat{\psi}_{\alpha }(\mathbf{r}) $ operators are components of the Nambu-spinor
\begin{align}\label{eq:NambuSpinor}
\Psi^{\dagger}(\mathbf{r}) = 
\left( 
\hat{\psi}^{\dagger}_{\uparrow }(\mathbf{r}), 
\hat{\psi}^{\dagger}_{\downarrow }(\mathbf{r}),
\hat{\psi}_{\uparrow }(\mathbf{r}),
\hat{\psi}_{\downarrow }(\mathbf{r}) 
\right).    
\end{align}
Later, we use the shorthand $\Psi^{\dagger}_{\alpha}(\mathbf{r}) = \left( \hat{\psi}^{\dagger}_{\alpha }(\mathbf{r}), \hat{\psi}_{\alpha }(\mathbf{r}) \right)$.

To simplify notation we have used the fact that the spin dependent interactions can be expressed in terms of the Pauli matrices as
\begin{align}\label{eq:spinchangeofbasis}
v^{\delta\gamma}_{\alpha\beta}(\mathbf{r},\mathbf{r}^{\prime})
=
\sum_{IJ} 
\sigma^{I}_{\alpha\delta}v^{IJ}(\mathbf{r},\mathbf{r}^{\prime})\sigma^{J}_{\beta\gamma},
\end{align}
where $\sigma^{i}$ is the Pauli matrix for $i=x,y,z$ and $\sigma^{0}$ is the $2\times 2$ identity matrix. Capital letters $I,J$ run over $0,x,y,z$, while Greek letters (e.g. $\alpha,\beta,\delta,\gamma$) take values $\pm 1$ enumerating the spin. The interaction $v^{IJ}$ may be grouped into three distinct types similar to those used by Bethe and Salpeter\cite{bethe2013quantum,itoh1965derivation}, (i) the Coulomb and the orbit-orbit interaction corresponding to the classical electromagnetic interaction of the electrons, 
\begin{align}
\sigma^{0}_{\alpha\delta}v^{00}(\mathbf{r},\mathbf{r}^{\prime})\sigma^{0}_{\beta\gamma},
\end{align}
(ii) the interaction between spin magnetic moments of the electrons, 
\begin{align}
\sigma^{i}_{\alpha\delta}v^{ij}(\mathbf{r},\mathbf{r}^{\prime})\sigma^{j}_{\beta\gamma},
\end{align}
and (iii) the spin-orbit magnetic coupling between electrons, 
\begin{align}
\sigma^{i}_{\alpha\delta}v^{i0}(\mathbf{r},\mathbf{r}^{\prime})\sigma^{0}_{\beta\gamma}.
\end{align}
Similarly, these spin-dependent interactions facilitate electron-nuclei coupling. This captures both weak and strong spin-orbit coupling regimes. 
The electron $n^{I}_{e}$ and nuclear $n^{I}_{n}$ densities are defined as:
\begin{align}
n^{I}_{e}(\mathbf{r}) =& \sum_{\alpha\beta} \hat{\psi}^{\dagger}_{\alpha}(\mathbf{r}) \sigma^{I}_{\alpha\beta} \hat{\psi}_{\beta}(\mathbf{r}),\\
n^{I}_{n}(\mathbf{r}\tau) =& \sum_{\kappa p} X^I_{\kappa}\delta(\mathbf{r}-\pmb{\tau}_{\kappa p}(\tau)),
\end{align}
with the total density given by $n^{I}=n^{I}_{e} + n^{I}_{n}$, where $n^{0}_{e}~(n^{0}_{n})$ is the electronic (nuclear) charge density and $n^{i}_{e}~(n^{i}_{n})$ gives the three components of the electronic (nuclear) spin density. Additionally, $X^{0}_{k}=-Z_{\kappa}$ and $X^{i}_{k}=S_{\kappa}$ are the atomic number and nuclear magnetic moment of nucleus $\kappa$, respectively, and $\pmb{\tau}_{\kappa p}(\tau)=\mathbf{R}_{p}+\pmb{\tau}_{\kappa}(\tau)$ is the time-dependent position of nucleus $\kappa$ in the crystal expressed as the location $\pmb{\tau}_{\kappa p}(\tau)$ in unit cell $p$ described by the lattice vectors $\mathbf{R}_{p}$. The infinitely extended solid is described using Born–von Karman boundary conditions, similar to Ref.~\onlinecite{giustino2017electron}, where periodic boundary conditions are applied to a large supercell which contains $N_{p}$ unit cells, described by the lattice vectors $\mathbf{R}_{p}$, with $p=1,\cdots,N_{p}$. For brevity we consider one species of nuclei, but this can be readily generalized following Ref.~\onlinecite{harkonen2024quantum}.
Finally, the external pairing fields $\Delta~(\bar{\Delta})$ are fully antisymmetric obeying 
\begin{subequations}
\begin{align}
\Delta_{\alpha\beta}(\mathbf{r},\mathbf{r}^\prime) &= -\Delta_{\beta\alpha}(\mathbf{r}^\prime,\mathbf{r}), \\
\mbox{and}\nonumber\\
\Delta^{*}_{\alpha\beta}(\mathbf{r},\mathbf{r}^\prime) &= \bar{\Delta}_{\alpha\beta}(\mathbf{r},\mathbf{r}^\prime),
\end{align}
\end{subequations}
and may be categorized into the various spin channels via the Balian-Werthamer matrices
\begin{align}
\Upsilon^I_{\alpha\beta}=\left[i\sigma^I\sigma^y\right]_{\alpha\beta},
\end{align} 
where the three matrices $\Upsilon^{x,y,z}$ form the symmetric (triplet) part of the spin component of the pairing function, whereas the antisymmetric (singlet) part is represented by the zeroth matrix $\Upsilon^{0}$. 

For this generalized Hamiltonian, we have derived the following closed set of Gor'kov-Hedin equations:
\begin{subequations}\label{eq:allHedin}
\begin{align}
&\mathbf{\Sigma}_{\eta\nu}(1,5)=-w^{LJ}(6,1)\pmb{\sigma}^{J}_{\eta\gamma} \pmb{\mathcal{G}}_{\gamma\mu}(1,4) \mathbf{\Lambda}^{L}_{\mu\nu}(4,5;6), \label{eq:hedineqSE}\\
&w^{LJ}(6,1)=w^{LJ}_{e}(6,1)+w^{LJ}_{ph}(6,1), \label{eq:HedinScreenedInt}\\
&w^{LJ}_{e}(6,1)=v^{LJ}(6,1)+v^{LM}(6,3)p^{MN}_{e}(3,4)w^{NJ}_{e}(4,1),\\
&w^{LJ}_{ph}(6,1)=w^{LM}_{e}(6,3)D^{MN}(3,4)w^{JN}_{e}(1,4),\label{eq:hedinwph}\\
&p^{MN}_{e}(7,8)=\left[\pmb{\mathcal{G}}_{\delta\mu}(7,9)
\mathbf{\Lambda}^{N}_{\mu\nu}(9,10;8)\pmb{\mathcal{G}}_{\nu\alpha}(10,7^+) \pmb{\sigma}^{M}_{\alpha\delta}\right]^{00},\\
&\mathbf{\Lambda}^{L}_{\mu\nu}(4,5;6)=
\delta(6,4)\delta(4,5)\pmb{\sigma}^{L}_{\mu\nu}\nonumber\\
&+\frac{\delta \mathbf{\Sigma}_{\mu\nu}(4,5) }{\delta \mathcal{G}^{ij}_{\alpha\beta}(9,10) }
\mathcal{G}^{im}_{\alpha\gamma}(9,11)
\Lambda^{L~mn}_{\gamma\eta}(11,12;6)
\mathcal{G}^{nj}_{\eta\beta}(12,10),\label{eq:hedinVertex}\\
&\pmb{\mathcal{G}}_{\eta \xi}(1,2)=\pmb{\mathcal{G}}_{H\eta\xi}(1,2)+\pmb{\mathcal{G}}_{H\eta\alpha}(1,3)\mathbf{\Sigma}_{\alpha\beta}(3,4)\pmb{\mathcal{G}}_{\beta\xi}(4,2).\label{eq:hedinDyson}
\end{align}
\end{subequations}
where the self-energy $\mathbf{\Sigma}$ is related to the Green's function $\pmb{\mathcal{G}}$  and the screened interaction $w$, using the electronic polarizability $p_e$, the nuclei fluctuation $D$, and the vertex function $\mathbf{\Lambda}$. The Dyson equation connects the full interacting Green's function to its non-interacting counterpart to close the set of equations. 

Similar to the two-particle interaction, Capital letters $L,J,M,K$ index the Pauli matrices running over $0,x,y,z$, while Greek letters enumerate the spin degrees of freedom taking values $\pm 1$. Moreover, $\pmb{\sigma}^{J}_{\eta\gamma}$ is the block Pauli matrix defined in Eq.~\ref{eq:blockpauli}. We have also introduced the short hand $(2) \equiv (\mathbf{x}_{2},\tau_{2}  )$, and used Einstein notation where repeated indices are summed and repeated variables represented by numbers are integrated over space-time, unless they appear on both sides of the equation.

Here, the bold letters are matrices in Nambu space, for example in the case of the self-energy and the vertex function, $\mathbf{\Sigma}_{\eta\nu}=\left[\Sigma^{mn}_{\eta\nu}\right]$ and $\mathbf{\Lambda}^{L}_{\mu\nu}=\left[ \Lambda^{Lmn}_{\mu\nu} \right]$, respectively, where the Lower case letters $m,n$ enumerate the Nambu components by taking values $0$ or $1$. Matrix multiplication is implied between two or more bold symbols. The polarizability $p_{e}$ and screened interactions $w$, $w_{e}$, and $w_{ph}$ are left not bold and lower case to indicate they are scalars in the Nambu space. We note the screened interaction and polarizability are scalars in Nambu space because interactions $\hat{\mathcal{U}}_{e-e}$ and $\hat{\mathcal{U}}_{e-n}$ conserve particle number. If these interactions did not conserve particle number, i.e., interactions that directly facilitate particle-condensate and/or condensate-condensate interactions, the screened interaction and polarizability would gain non-trivial Nambu matrix elements. Consequently, it is sufficient to use auxiliary electric and magnetic perturbing fields to reduce the two-particle Green's functions similar to Hedin's original work, see Appendix~\ref{Appendix:derivationGH}. We note that here the pairing field $\Delta$ is explicitly included as a {\it physical} field and is not necessary to construct the Gor'kov-Hedin equations since our lowest-order vertex approximation yields the GW approximation. This is in contrast to Ref.~\onlinecite{marie2024anomalous} where {\it auxiliary} pairing fields are needed to write Eqs. (\ref{eq:hedineqSE}) - (\ref{eq:hedinDyson}) such that the T-matrix emerges naturally at lowest-order.

The matrix Nambu Green's function\cite{nambu1960quasi} is given by 
\begin{align}
\pmb{\mathcal{G}}_{\eta\xi}(1,2) &=
\left[\begin{array}{cc}
G_{\eta\xi}(1,2) &F_{\eta\xi}(1,2)\\
-\bar{F}_{\eta\xi}(1,2)& -\bar{G}_{\eta\xi}(1,2) 
\end{array}\right],
\end{align}
where $G,\bar{G}$ are ordinary and $F,\bar{F}$ are anomalous single-particle Green's functions, respectively, defined by the components of imaginary-time-ordered matrix propagator 
$\mathfrak{G}_{\eta\xi}(1,2)=-\braket{ \mathcal{T} \{ \Psi_{\eta}(1)\otimes\Psi^{\dagger}_{\xi}(2) \} }$, see Appendix~\ref{Appendix:derivationGH} for details. The off diagonal terms determine the anomalous density, in analogy to $G~(\bar{G})$ giving the normal density, and are zero for a non-superconducting system. If the eigenstates in the thermal ensemble average correspond to a fixed particle number $N$, the off diagonal terms are strictly zero, however, if the eigenstates have the form of a BCS wave function a non-trivial result is possible.

The self-energy $\mathbf{\Sigma}$ is a block matrix similar to the Nambu Green's function where the diagonal blocks are the ordinary components that describe the electron and phonon exchange-correlation effects that give way to mass renormalizations and quasiparticle lifetimes, while the off-diagonal blocks of $\mathbf{\Sigma}$ yield the superconducting spectral gap function. The self energy naturally partitions into electronic and phonon contributions: $\mathbf{\Sigma}_{\eta\nu}(1,5)=\mathbf{\Sigma}_{\eta\nu}^{elec}(1,5)+\mathbf{\Sigma}_{\eta\nu}^{ph}(1,5)$, due to the additive nature of the screen interaction, although these two are inextricably linked through the vertex. All this makes $\mathbf{\Sigma}$ a convenient window into how the mutual interactions between electrons and phonons modify the electronic (phonon) properties and promote or suppress superconductivity.

When the two-particle Green's function is expressed in terms of $\pmb{\mathcal{G}}$ and its functional derivative [Eq.~\ref{eq:trick}], the resulting generalized Hartree potential $V_{H}^{J}(1)\equiv\braket{\mathcal{T}\{ n^{I}(3)\} } v^{IJ}(3,1)$ is a local multiplicative factor that maybe combined with the free Nambu Green's function, to define the Hartree propagator:
\begin{align}\label{eq:hartree_prop}
\pmb{\mathcal{G}}^{-1}_{H\eta\alpha}(1,3)=\pmb{\mathcal{G}}^{-1}_{0\eta\alpha}(1,3)-V_{H}^{J}(1)\pmb{\sigma}^{J}_{\eta\gamma}\delta(1,3),
\end{align}
where $\pmb{\mathcal{G}}^{-1}_{0\eta\alpha}(1,3)$ is the free Green's function:
\begin{align}
\begin{bmatrix}
-\frac{d}{d\tau_{1}}\delta_{\eta\beta}\delta(1,3)-h_{\eta\beta}(1)\delta(1,3), \quad -\bar{\Delta}_{\eta\beta}(1,3)\\
-\Delta_{\eta\beta}(1,3), \quad +\frac{d}{d\tau_{1}}\delta_{\eta\beta}\delta(1,3)-h^{*}_{\eta\beta}(1)\delta(1,3)
\end{bmatrix}.
\end{align}

Lastly, the screened interaction due to the lattice vibrations may be written in the Harmonic approximation by expanding $D$ to second order in nuclear displacements as:
\begin{align}\label{eq:wph_HA}
w_{ph}^{LJ}(6,1)&=
w_{e}^{LM}(6,7) 
\braket{\mathcal{T}\{ n^{M}_n(\mathbf{r}_7) n^{N}_n(\mathbf{r}_8) \} }
w^{JN}_{e}(1,8)  \nonumber\\
&+
g^{Li}_{\kappa p}(6,\tau_7) 
\mathfrak{D}^{ij}_{\kappa p,\kappa^\prime p^\prime}(\tau_7,\tau_8)
g^{Jj}_{\kappa^\prime p^\prime}(\tau_8,1)
\end{align}
where $g^{Li}_{\kappa p}(5,\tau_7)=w_{e}^{LM}(5,\mathbf{r}_7\tau_7) X^M_{\kappa}\nabla^i_{\mathbf{r}_7}\delta(\mathbf{r}_7-\pmb{\tau}_{\kappa p}^{0}) $ is the spin-dependent electron-phonon coupling potential, $\pmb{\tau}_{\kappa p}^{0}$ is the equilibrium position of nucleus $\kappa$ in unit cell $p$, $n^{M}_n(\mathbf{r}_7)~(n^{N}_n(\mathbf{r}_8))$ is the equilibrium nuclear density, and $\mathfrak{D}^{ij}_{\kappa p ,\kappa^\prime p^\prime}(\tau_7,\tau_8)$ is the the phonon Green's function. $\mathfrak{D}^{ij}_{\kappa p ,\kappa^\prime p^\prime}(\tau_1,\tau_2)$ satisfies the equation of motion:
\begin{align}
&M_{\kappa}\frac{d^2}{d\tau^2_1}\mathfrak{D}^{ij}_{\kappa p ,\kappa^\prime p^\prime}(\tau_1,\tau_2)=\delta_{\kappa p ,\kappa^\prime p^\prime}\delta_{ij}\delta(\tau_1,\tau_2)\nonumber\\
&+\sum_{\kappa^{\prime\prime} p^{\prime\prime}l}\int \mathrm{d}\tau^\prime ~\Pi^{il}_{\kappa p,\kappa^{\prime\prime} p^{\prime\prime}}(\tau_1,\tau^\prime)
\mathfrak{D}^{lj}_{\kappa^{\prime\prime} p^{\prime\prime} ,\kappa^\prime p^\prime}(\tau^\prime,\tau_2),\label{eq:baym}
\end{align}
where
\begin{align}\label{eq:SEphonon}
&\Pi^{il}_{\kappa p,\kappa^{\prime\prime} p^{\prime\prime}}(\tau_1,\tau^\prime)=\nonumber\\
&\iint \mathrm{d}\mathbf{r}\mathrm{d}\mathbf{r}^{\prime} 
[X^I_{\kappa}\nabla^i_{\mathbf{r}}\delta(\mathbf{r}_1-\pmb{\tau}_{\kappa p}^{0})
    w_{e}^{IJ}(\mathbf{r}\tau,\mathbf{r}^\prime \tau^\prime) X^J_{\kappa^\prime}\nabla^l_{\mathbf{r}^\prime}\delta(\mathbf{r}^\prime-\pmb{\tau}_{\kappa^{\prime\prime} p^{\prime\prime}}^{0})\nonumber\\
    &-\delta_{\kappa p,\kappa^{\prime\prime} p^{\prime\prime}}\delta(\tau,\tau^\prime)\nabla^i_{\mathbf{r}}\braket{n^{I}(\mathbf{r})}v^{IJ}(\mathbf{r},\mathbf{r}^\prime)X^J_{\kappa^\prime}\nabla^l_{\mathbf{r}^\prime}\delta(\mathbf{r}^\prime-\pmb{\tau}_{\kappa^{\prime\prime} p^{\prime\prime}}^{0})]
\end{align}
is the phonon self-energy, see Appendix~\ref{appendix:Phonon} for details. By inspection, $\Pi$ is nearly the same as presented in Ref.~\onlinecite{giustino2017electron}, where the first term describes the action of the electrons on the lattice vibrations in the crystal and the second term is the static force experienced by the nuclei in their equilibrium configuration. Here, in addition to the Coulomb interaction between nuclei their mutual spin-spin and spin-orbit magnetic couplings are also included. A previous work that generalized Hedin's equations to a system of superconducting electrons coupled with a system of phonons critically neglect the feedback between lattice and electrons along with any spin dependence\cite{linscheid2015hedin}. 

We note that the nuclei affect the electronic structure via the dielectric matrix which enters $w_{ph}$ in Eq.~\ref{eq:hedinwph} and through the nuclear density inside the Hartree potential. Similarly, according to Eq.~\ref{eq:SEphonon}, the coupling of the nuclear displacements to the electrons is completely defined by the electronic dielectric matrix through $w_e$. Thus, the dielectric matrix $\varepsilon^{IJ}_{e}$
plays a central role in coupling the electronic and phonon systems.

The complete set self-consistent Gor'kov-Hedin equations for the electrons combined with those for the nuclei, derived first by Baym (Ref.~\onlinecite{baym1961field}), may be referred to as the Gor'kov-Hedin-Baym equations. Apart from invoking the harmonic approximation, the Gor'kov-Hedin-Baym equations describe the the coupled electron-phonon system entirely from first principles and form a highly sophisticated description of interacting electrons and phonons in the presence of superconductivity. 

\begin{figure}[t]
\begin{center}
\includegraphics[width=1.0\columnwidth]{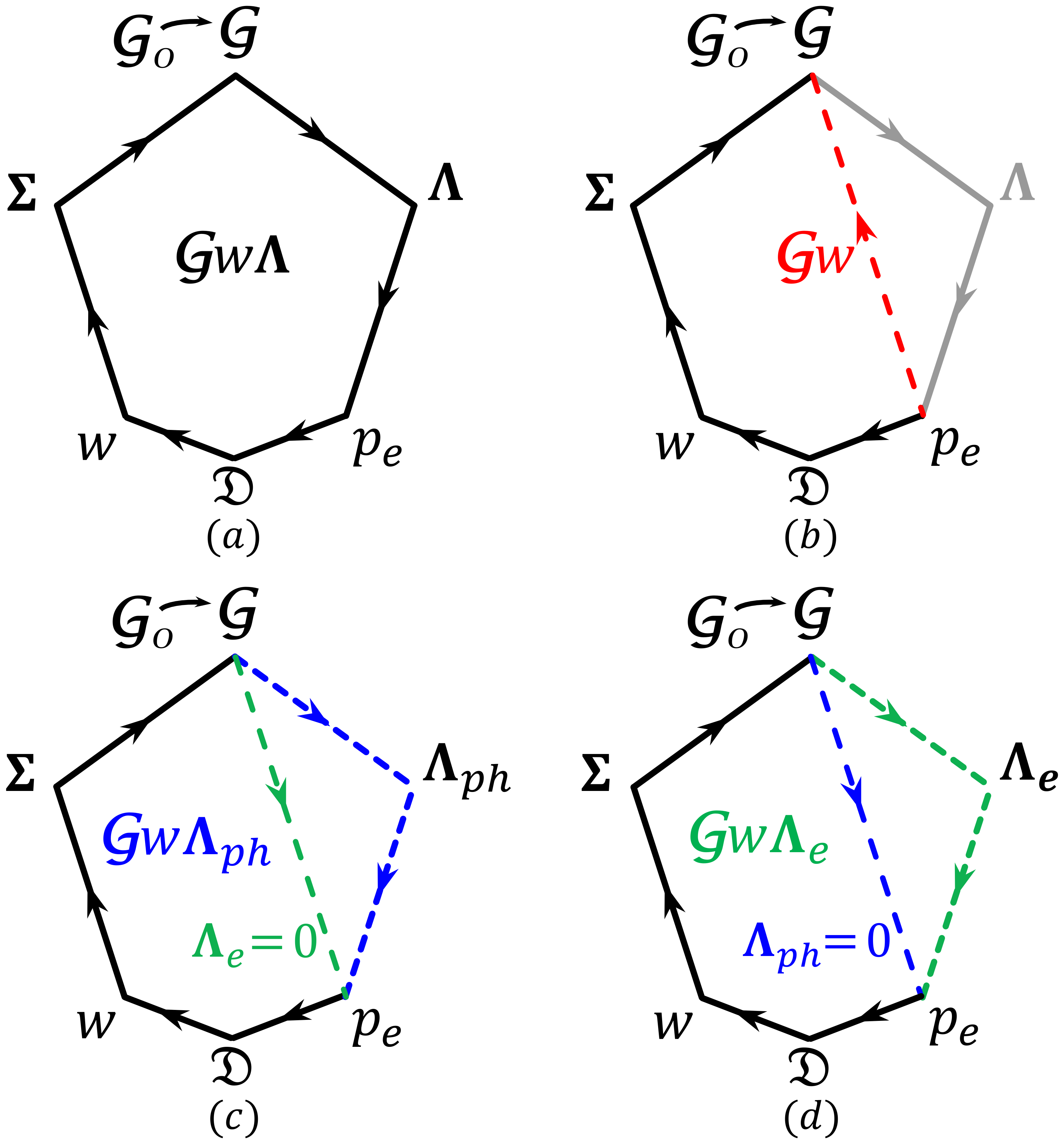}
\caption{(color online) Schematic of self-consistent cycle for (a) the full set of Gor'kov-Hedin equations, (b) the GW approximation, and where the vertex is restricted to (c) phonon or (d) electron degrees of freedom.} 
\label{fig:HEDINLOOP} 
\end{center}
\end{figure}

\section{Iterating the Gor'kov-Hedin-Baym Equations}\label{sec:iterating}
The Gor'kov-Hedin-Baym equations form a set of equations that must be solved self-consistently. This means that the Green's function that enters into the self-energy is necessarily the same as that in Dyson's equation with the same self-energy. This is quite a daunting task in general. In practice one must find a simplification of the Gor'kov-Hedin-Baym equations. Since there is no straightforward manner by which to define a convergent perturbation expansion, due to the lack of a small parameter, the approach to simplifying this set of equations is a bit arbitrary and needs careful analysis. 

For Hedin's original equations, self-consistency is formally approached via systematic iteration of the equations\cite{schindlmayr1998systematic}. Analogously for the Gor'kov-Hedin-Baym equations this is accomplished by starting with a trial Nambu Green's function, generally the free propagator $\mathcal{G}_0$, and then constructing $\Lambda$, $p_e$, $D$, $w$, and $\Sigma$ in succession. To close the loop,  Dyson's equation is used to obtain a new trial Nambu Green's function. This process is repeated until the full set of Gor'kov-Hedin-Baym equations are satisfied simultaneously yielding the exact electron (normal and anomalous) and phonon Green's functions, as illustrated in Fig.~\ref{fig:HEDINLOOP}(a). Since we are working in the Harmonic approximation, no extra self-consistency loops are needed to determine $\mathfrak{D}$. If one wishes to include anharmonic effects (e.g. phonon-phonon interactions) the resulting mathematical formalism is significantly more complex, thereby requiring further iteration cycles for the fully interacting phonon propagator\cite{gillis1970self,marini2018functional,harkonen2020many}. Practically, full self-consistency is presently out of reach for almost all material calculations, compelling us to employ approximate partial self-consistency schemes, e.g., Fig.~\ref{fig:HEDINLOOP}(b)-(d). 

In the following sections we explicitly iterate the Gor'kov-Hedin-Baym equations to examine the corrections to the free electron and phonon propagators that arise and the physical processes they represent, particularly, in regard to the role superconductivity plays.    

\subsection{Non-Interacting and Hartree Approximations}
Before iterating the Gor'kov-Hedin-Baym equations we start by analyzing the non-interacting and Hartree Nambu Green’s function. By inspection, we recognize that $\mathcal{G}_{0}$ can be written as  
\begin{align}
\pmb{\mathcal{G}}^{-1}_{0\eta\alpha}(1,3)=\tau^z \left[ -\frac{d}{d\tau_{1}}\delta_{\eta\beta}\delta(1,3)\tau^0-
\mathcal{H}_{BdG}(1,3)\right]
\end{align}
where $\tau^{z}$ is equivalent to $\sigma^{z}$ except it acts on the Nambu indices, the matrix in the brackets is the Abrikosov-Gor'kov propagator, and
\begin{align}
\mathcal{H}_{BdG}(1,3)=
\begin{bmatrix}
h_{\eta\beta}(1)\delta(1,3) & \bar{\Delta}_{\eta\beta}(1,3)\\
-\Delta_{\eta\beta}(1,3) & -h^{*}_{\eta\beta}(1)\delta(1,3)
\end{bmatrix}
\end{align}
is the Bogoliubov-de Gennes (BdG) Hamiltonian\cite{sigrist1991phenomenological,zhu2016bogoliubov}. The BdG Hamiltonian typically appears as a result of a mean-field treatment and provides a general quadratic Hamiltonian that describes the quasiparticle excitations in superconducting systems. Here, it appears as a starting point for further self-energy corrections. The BdG Hamiltonian matrix has particle-hole symmetry, which implies an associated redundancy in the Hamiltonian. Specifically, the components of the Nambu spinor in Eq.~\ref{eq:NambuSpinor} are related by the complex conjugate, and hence they are not independent. Accordingly,
the matrix components of $\mathcal{H}_{BdG}$ are not independent of each other and equivalently, the eigenvalue spectrum of $\mathcal{H}_{BdG}$ 
come in pairs\cite{sato2017topological,sigrist1991phenomenological,zhu2016bogoliubov}. 

In systems where spin-orbit coupling can be neglected, the superconducting gap $\Delta_{\eta\beta}(1,3)$ may be classified by its spin state (singlet or triplet) and its total angular momentum, e.g., $s$, $p$, $d$, $f$, and so on. In the presence of spin-orbit coupling this scheme breaks down since spin and angular momentum are not good quantum numbers. In such cases, the pair potential is classified by using the crystal symmetries\cite{sigrist1991phenomenological,sato2017topological}. In particular, special attention is often paid to inversion symmetry since the pair potential in a centrosymmetric material has a definite parity under inversion, either positive or negative. In the case of noncentrosymmetric systems, inversion is broken intrinsically and parity mixing occurs, i.e., the coexistence of spin-singlet and spin-triplet pair potentials\cite{frigeri2004superconductivity,bauer2012non}. Finally, if the systems exhibits magnetic ordering as is the case for superconductivity in altermagnets\cite{mazin2203notes}, the magnetic and spin space group must also be considered\cite{feng2025superconducting}. Since $\Delta$ is included here as an external pairing field it need not obey or be restricted by the same symmetries observed by $\hat{\mathcal{T}}_{e}$.

\begin{figure}[t]
\begin{center}
\includegraphics[width=0.99\columnwidth]{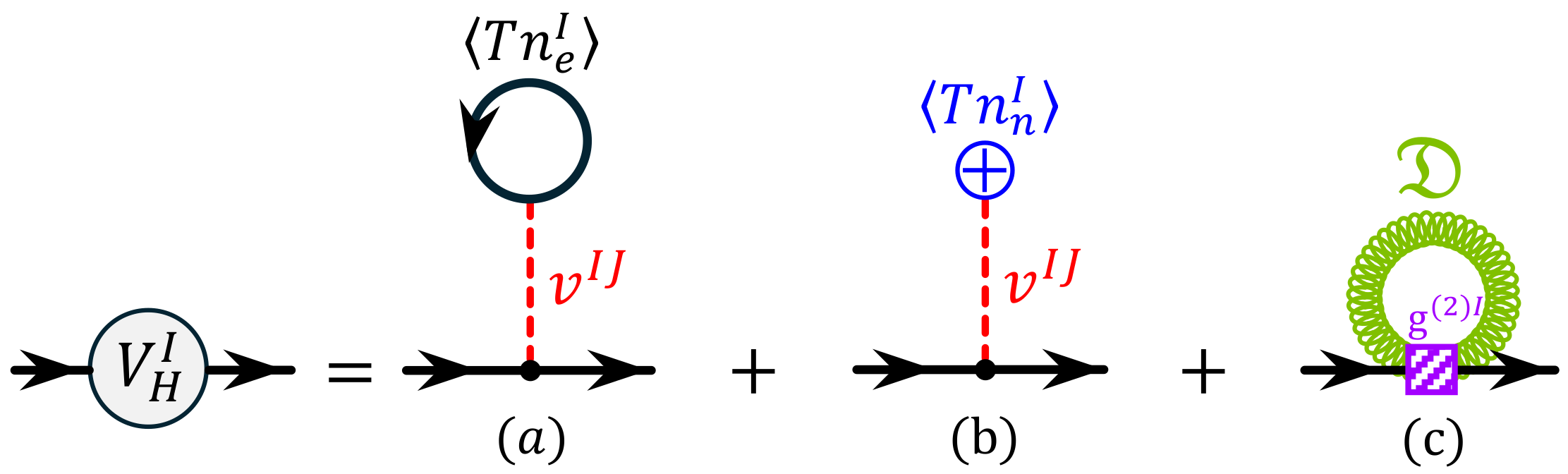}
\caption{ (color online) Diagrammatic representation of the various contributions to the Hartree potential in the harmonic approximation. (a) electronic Hartree self-energy, (b) clamped-ion potential, and (c) Debye-Waller self-energy.} 
\label{fig:Hartree} 
\end{center}
\end{figure}

The zeroth order correction to the free Nambu Green’s function is the Hartree potential. To make the physical content of the Hartree potential clear the expectation of the nuclear density is expanded in nuclear displacements and truncated at the second order in accord with the harmonic approximation, yielding:
\begin{align}
V_{H}^{J}(1)&\approx\braket{\mathcal{T}\{ n^{I}_e(2)\} } v^{IJ}(2,1)\nonumber\\
&+
\braket{\mathcal{T}\{ n^{I}_n(\mathbf{r}_2)\} } v^{IJ}(\mathbf{r}_2,\mathbf{r}_1)\delta(\tau_2,\tau_1)\nonumber\\
&-
\sum_{\kappa p}\mathfrak{D}^{ij}_{\kappa p,\kappa p}(\tau_2,\tau_2)g^{(2)Jij}_{\kappa p}(\mathbf{r}_1)\delta(\tau_2,\tau_1)
\end{align}
where $g^{(2)Jij}_{\kappa p}(\mathbf{r}_1)=\frac{1}{2}X^I_{\kappa p}\nabla^i_{\mathbf{r}_2}\nabla^j_{\mathbf{r}_2}\delta(\mathbf{r}_2-\pmb{\tau}^0_{\kappa p})v^{IJ}(\mathbf{r}_2,\mathbf{r}_1)$ is the spin-dependent second-order electron-phonon coupling potential. The first term is the well known Hartree contribution that is present in Hedin's original work, the second term is the clamped-ion potential that is common in first-principles electronic structure calculations where the nuclei are described as classical particles clamped in their equilibrium positions, and the third term is the Debye-Waller self-energy which captures the time-dependent fluctuations of the nuclear density around the nuclear equilibrium positions. A diagrammatic representation of $V_{H}^{J}$ in the harmonic approximation is given in Fig.~\ref{fig:Hartree}.

In addition to the Coulomb interaction between an electron and the average density of electrons in the system, and the static classical and quantum correlations of the equilibrium nuclear configuration of the system, the Hartree potential also captures the spin-spin and spin-orbit magnetic coupling between an electron and the electronic magnetic density and static nuclear polarizations that come about from the spin texture of the system. The spin dependence of $V_{H}^{J}$ induces spin-flip matrix elements in $\mathcal{G}_{H}$ that can influence the solution of the self-consistent Gor'kov-Hedin-Baym equations.

The Hartree propagator can then be understood by iterating the Dyson-like equation in Eq.~\ref{eq:hartree_prop}. Interestingly, $V_{H}^{J}$ only couples to diagonal Nambu matrix elements which stems from writing the two-particle Green's function in terms of the functional derivative with respect to an auxiliary electromagnetic field [Eq.~\ref{eq:trick}]. Consequently, the Hartree self-energy {\it implicitly} impacts the superconducting state via the density in the electronic Hartree and Debye-Waller terms in contrast to the typical BCS mean-field equations that are explicitly decoupled in the pairing channel\cite{gor1958energy,gor1959microscopic,sigrist1991phenomenological,zhu2016bogoliubov}. 

\subsection{First Iteration: The GW Approximation}\label{sec:first_iteration}

Starting from the Hartree Green's function, the first iteration of the Gor'kov-Hedin-Baym equations yields a vertex function that is diagonal in space-time and Nambu degrees of freedom,
\begin{align}
\mathbf{\Lambda}^{L}_{\mu\nu}(4,5;6)=
\delta(6,4)\delta(4,5)\pmb{\sigma}^{L}_{\mu\nu}.
\end{align}
Consequently, the electronic polarization reduces to its non-interacting form:
\begin{align}\label{eq:polGWA}
p^{MN}_{e}(7,8)=\left[
\pmb{\mathcal{G}}_{\delta\mu}(7,8)
\pmb{\sigma}^{N}_{\mu\nu}
\pmb{\mathcal{G}}_{\nu\alpha}(8,7^+) 
\pmb{\sigma}^{M}_{\alpha\delta}\right]^{00},
\end{align}
and the self-energy is then
\begin{align}\label{eq:eGWA}
&\mathbf{\Sigma}_{\eta\nu}^{GW}(1,5)=-
w^{LJ}(5,1)\pmb{\sigma}^{J}_{\eta\gamma} \pmb{\mathcal{G}}_{\gamma\mu}(1,5) \pmb{\sigma}^{L}_{\mu\nu},
\end{align}
the so-called GW approximation (GWA). 

\subsubsection{RPA Polarizability and Screened Interaction}\label{sec:polarizeability}
\begin{figure}[t!]
\begin{center}
\includegraphics[width=0.99\columnwidth]{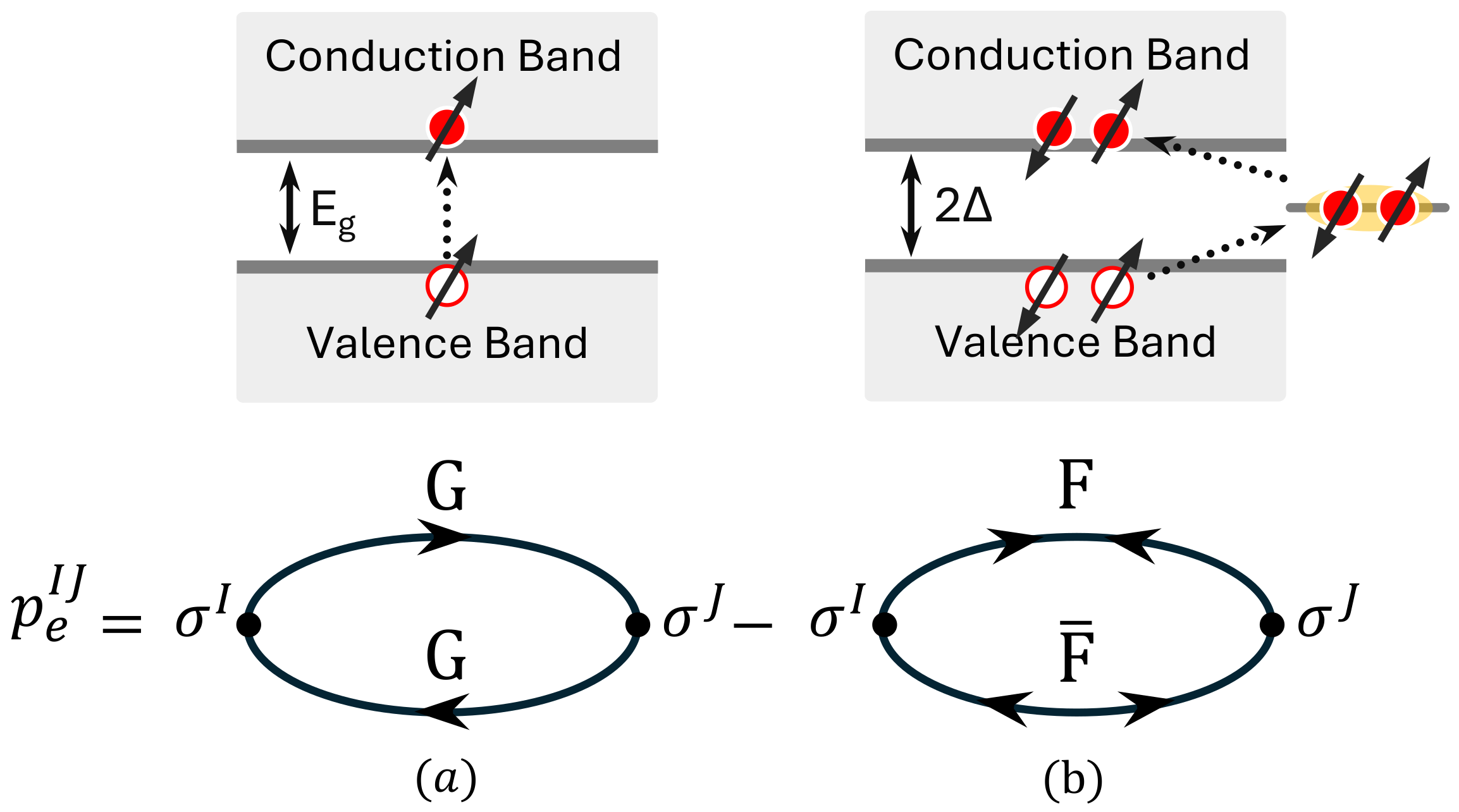}
\caption{ (color online) Diagrammatic representation of the (a) electron-hole and (b) Cooper pair breaking excitations within the RPA that contribute to the electronic polarizability. A schematic of the electronic band diagrams of a (a) semiconductor and a (b) superconductor are shown above to illustrate each associated excitation process. } 
\label{fig:excitation} 
\end{center}
\end{figure}

The first iteration of the Gor'kov-Hedin-Baym equations yields the non-interacting irreducible polarizability:
\begin{align}\label{eq:polGWA_unroll}
p^{MN}_{e}(7,8)&=
G_{\delta\mu}(7,8)\sigma^{N}_{\mu\nu}G_{\delta\mu}(8,7^+)\sigma^{M}_{\alpha\delta}\nonumber\\
&-
F_{\delta\mu}(7,8)\sigma^{N}_{\mu\nu}\bar{F}_{\delta\mu}(8,7^+)\sigma^{M}_{\alpha\delta}.
\end{align}
In the normal state, only the first term is present and is a direct generalization of the standard charge polarizability discussed by Lindhard\cite{lindhard1954properties,mahan2013many} and Hedin\cite{hedin1965}. This polarization `bubble' describes the spontaneous creation and annihilation of particle-hole pairs from the vacuum whose spin structure is a direct consequence of the underlying single-particle Green's function, as illustrated in Fig.~\ref{fig:excitation}(a). 

In the superconducting state, the second term in Eq.~\ref{eq:polGWA_unroll} is finite, albeit typically weaker than the ordinary term since $F(\bar{F})$ is proportional to the the anomalous spectral gap. The anomalous polarizability describes an effective particle-hole fluctuation channel via the spontaneous excitation of two electrons out of the condensate while two free electrons simultaneously join the condensate. The added electrons and holes left behind mirror that of an electron-hole excitation process, as illustrated in Fig.~\ref{fig:excitation}(b). 

The spin-dependence of these fluctuations, and associated superconducting paring symmetry embeded in $F(\bar{F})$, may be measured in nuclear magnetic resonance (NMR) spectroscopy where a shift in the resonance frequency of a nucleus upon entering the superconducting state is caused by the internal magnetic structure of the Cooper pairs, the so-called Knight shift\cite{abrikosov1962spin,yosida1958paramagnetic,anderson1959knight}. Thus, providing a direct way to distinguish between even and odd-parity pairing solutions.

Following the Gor'kov-Hedin-Baym equations [Fig.~\ref{fig:HEDINLOOP}], $w_e$, $w_{ph}$, and $\mathfrak{D}$ are constructed from $p_e$ through the inverse dielectric matrix. For the normal state, if $\varepsilon^{IJ}_e$ is large as expected for metals and narrow band gap materials, then $w_e$ is small. This suggests the electron-phonon coupling $g^{Li}_{\kappa p}$ and the electronic corrections to $\Pi$ are relatively weak. In contrast, if the system is gapped, $w_e$ is large, yielding strong electron-phonon coupling, phonon propagator self-energy corrections, and electronic structure renormalizations. The sensitivity of the lattice to the screening environment provides a powerful probe of phase transitions, where as a system crosses from a metallic normal state to a gapped superconducting phase, changes in the elastic tensor (as a result of changes in $\varepsilon^{IJ}_e$) can be directly measured and tracked across the phase transition using resonant ultrasound spectroscopy\cite{balakirev2019resonant,migliori1990complete,migliori1990elastic,ghosh2021thermodynamic}. Furthermore, it gives rise to a delicate feedback between the electrons and phonons which may enhance superconducting critical temperatures\cite{zhang2025magnetism,moghadas2025effective,abramovitch2025electron}.

\subsubsection{The GW Self-Energy}\label{sec:gwa}

\begin{figure}[t]
\begin{center}
\includegraphics[width=0.99\columnwidth]{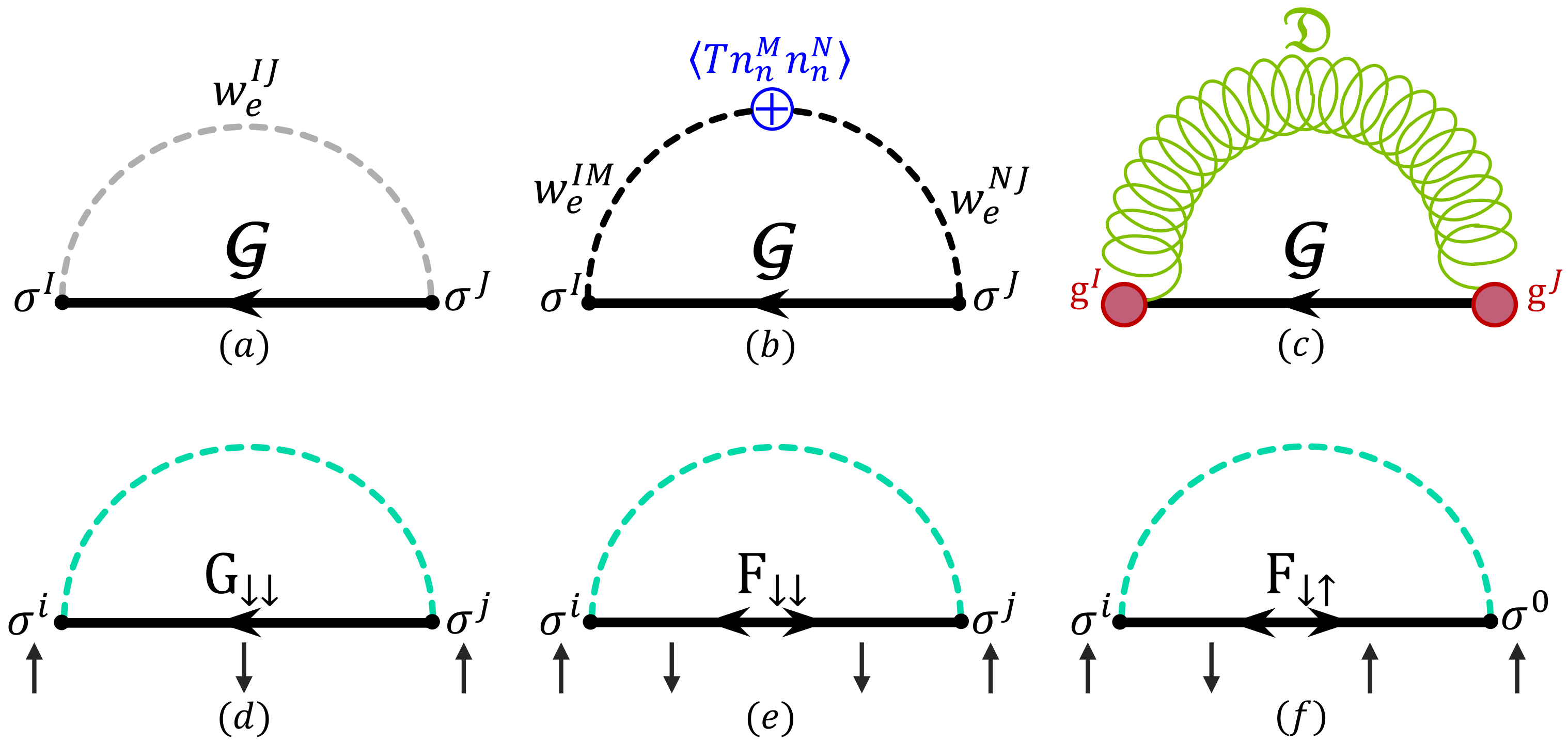}
\caption{(color online) Diagrammatic representation of the GW self-energy: (a) electronic, and (b) clamped-ion and (c) Fan-Migdal phonon contributions to the self-energ. (d)-(f) Illustrate the effect of spin-dependent interactions on a spin up electron for both (d) conventional and (e)-(f) anomalous components of the self-energy (see text for details). The dashed teal lines represent either the electronic $(w_{e})$ or phonon $(w_{ph})$ screened interactions.} 
\label{fig:SE} 
\end{center}
\end{figure}

The GW self-energy describes spin-dependent screened exchange processes arising from the electrons:
\begin{align}\label{eq:eGWA_electronic}
&\mathbf{\Sigma}_{\eta\nu}^{GWe}(1,5)=-
w^{LJ}_e(5,1)\pmb{\sigma}^{J}_{\eta\gamma} \pmb{\mathcal{G}}_{\gamma\mu}(1,5) \pmb{\sigma}^{L}_{\mu\nu},
\end{align}
and the lattice,
\begin{align}\label{eq:eGWA_phonon_HA}
&\mathbf{\Sigma}_{\eta\nu}^{GWph}(1,5)=\nonumber\\
&-
w_{e}^{LM}(5,6) 
\braket{\mathcal{T}\{ n^{M}_n(\mathbf{r}_6) n^{N}_n(\mathbf{r}_7) \} }
w^{JN}_{e}(1,7)\pmb{\sigma}^{J}_{\eta\gamma} \pmb{\mathcal{G}}_{\gamma\mu}(1,5) \pmb{\sigma}^{L}_{\mu\nu}  \nonumber\\
&-
g^{Li}_{\kappa p}(5,\tau_6) 
\mathfrak{D}^{ij}_{\kappa p,\kappa^\prime p^\prime}(\tau_6,\tau_7)
g^{Jj}_{\kappa^\prime p^\prime}(\tau_7,1)\pmb{\sigma}^{J}_{\eta\gamma} \pmb{\mathcal{G}}_{\gamma\mu}(1,5) \pmb{\sigma}^{L}_{\mu\nu},
\end{align}
where $\mathbf{\Sigma}^{GWph}$ is composed of the clamped-ion and the Fan-Migdal self-energies that come from the static and dynamic lattice correlations, respectively. A diagrammatic representation of the various components of the GW self-energy is given in Fig.~\ref{fig:SE}(a)-(c). In the absence of superconductivity, the GWA has been successful in providing a reasonable description of the screening environment in solids, yielding single-particle mass and band gap renormalizations and lifetime effects \cite{aryasetiawan1998gw,onida2002electronic,reining2018gw}. 

Importantly, the GWA lays the foundation for the Migdal-Eliashberg theory of conventional superconductors, wherein phonons are the key driver of electron-electron pairing\cite{allen1983theory,eliashberg1960interactions,eliashberg1961temperature,migdal1958interaction}. For these systems Migdal showed that corrections beyond the bare electron-phonon vertex are typically small in Fermi liquids\cite{migdal1958interaction}, since the electronic mass $m$ is significantly smaller than the ionic masses in the solid. Moreover, since electron correlations are considered minimal in these systems it is routine to neglect the electronic vertex contribution also and prioritize the description of the screening environment\cite{allen1983theory}. 

The Gor'kov-Hedin-Baym GW self-energy presented here is a generalization of the Migdal-Eliashberg theory to systems with spin-dependent interactions. To examine the physical meaning of these expressions, we first consider the well studied special case of a pure Coulomb interaction $\left( v^{IJ}=v^{00}\delta_{IJ}\delta_{I0} \right)$ and spin-independent nuclei fluctuations $\left(D^{MN}=D^{00}\delta_{MN}\delta_{N0}\right)$. Since the interaction is purely Coulombic, only the charge component of $w_{e}$ and $w_{ph}$ remain:
\begin{align}
w^{00}_{e}(6,1)&=v^{00}(6,1)+v^{00}(6,3)p^{00}_{e}(3,4)w^{00}_{e}(4,1)\\
w_{ph}^{00}(6,1)&=
w_{e}^{00}(6,7) 
\braket{\mathcal{T}\{ n^{0}_n(\mathbf{r}_7) n^{0}_n(\mathbf{r}_8) \} }
w^{00}_{e}(1,8)  \nonumber\\
&+
\hat{g}^{0i}_{\kappa p}(6,\tau_7) 
\mathfrak{D}^{ij}_{\kappa p,\kappa^\prime p^\prime}(\tau_7,\tau_8)
\hat{g}^{0j}_{\kappa^\prime p^\prime}(\tau_8,1)
\end{align} 
along with the charge channel of $p_e$,
\begin{align}
p^{00}_{e}(7,8)=\left[
\pmb{\mathcal{G}}_{\delta\mu}(7,8)
\pmb{\sigma}^{0}_{\mu\nu}
\pmb{\mathcal{G}}_{\nu\alpha}(8,7^+) 
\pmb{\sigma}^{0}_{\alpha\delta}\right]^{00}.
\end{align}
If the Nambu Green's function is diagonal in spin space, we recover the self-energy graphs  given by Hedin\cite{hedin1965}, Lundqvist and Hedin\cite{hedin1970effects}, Baym\cite{baym1962self,baym1961field}, Kadanoff and Baym\cite{kadanoff_quantum_1962,baym1961conservation}, and Giustino \cite{giustino2017electron} for the normal state. In the superconducting state, $\mathbf{\Sigma}^{GW}$ gives the self-energy graphs retained by Allen and Mitrovic\cite{allen1983theory}, and Scalapino\cite{scalapino1966strong} where phonon and Coulomb interactions are treated on the same footing within the Migdal-Eliashberg GW approximation.

However, if the electrons in the material system possess an inherent spin structure, the dependence of the self-energy on the spin degrees of freedom arises entirely from the Green function rather than the screened interaction, as similarly pointed out in Ref.~\onlinecite{aryasetiawan2008}. Furthermore, if a finite external pairing field $\Delta$ is present, then the self-energy gives way to non-zero anomalous off-diagonal components whose spin structure originates entirely from pairing field. This case is a generalization of the original Migdal-Eliashberg theory to spin-dependent Green's function and self-energy with purely Coulombic interaction. Here, it naturally emerges from the present formulation as a special case where spin interactions are absent. 

Now let us consider the case where the interactions between particles are spin dependent, which may arise from spin–spin or spin–orbit magnetic couplings. If a particle of spin $\uparrow$ enters an ordinary component of the self-energy its spin is flipped by the spin operator $\sigma_{\uparrow\downarrow}^{i}$ and (i) a magnon given by $w^{ij}_e$ is emitted via $\Sigma^{elec}$, and (ii) a spin-dependent phonon given by $w_{ph}^{ij}$ is excited via $\Sigma^{ph}$. Then upon exiting the self-energy the magnon (phonon) is reabsorbed, thereby flipping the spin by $\sigma_{\downarrow\uparrow}^{j}$, and recovering its original spin state. This process is illustrated in Fig.~\ref{fig:SE} (d). This is analogous to Eliashberg's original theory where an electron emits and absorbs a phonon without the possibility of a spin flip. 

Since the anomalous components of the self-energy describe the superconducting spectral gap and pairing symmetry of the superconducting state, the presence of spin-dependent interactions play an influential role in fostering Cooper pairs of various spin structures and nontrivial forms of superconductivity. If two electrons enter the superconducting condensate $F_{\downarrow\downarrow}$, one has their spin flipped by the spin operator $\sigma_{\uparrow\downarrow}^{i}$ by emitting a magnon $w^{ij}_e$ (phonon $w_{ph}^{ij}$), while the other has their spin flipped by absorbing a magnon (phonon). This yields a spectral gap in the opposite spin channel compared to the superconducting condensate [Fig.~\ref{fig:SE} (e)]. This cross-talk between spin channels is more apparent in the presence of spin–orbit coupling. That is, when two electrons enter the superconducting condensate $F_{\downarrow\uparrow}$ one emits a magnon (`spin' phonon) and the other absorbs a plasmon (`charge' phonon) via $w^{i0}_e$ ($w_{ph}^{i0}$), producing a spectral gap in the triplet channel despite $F_{\downarrow\uparrow}$ being in a singlet state, as illustrated in Fig.~\ref{fig:SE} (f). Therefore, by iterating the Gor'kov-Hedin-Baym equations a nontrivial superconducting state may emerge from a singlet pairing field due to the existence of relativistic effects. Reports of similar effects have been observed via proximity interactions\cite{trang2020conversion,sharma2022comprehensive}.

\subsection{Iterating Beyond the GWA: Fluctuations, Vertex Corrections, and Effective Quasiparticle Interactions}\label{sec:beyondgw}

A common thread that stitches unconventional superconductors (e.g., the cuprates, the Fe-pnictides, and heavy-fermion systems) together is the critical role fluctuations play in the pairing interaction. Unlike the conventional superconductors which follow the BCS theory, unconventional superconductors display behaviors that cannot be accounted for within this picture, thereby requiring contributions to the self-energy that are beyond the mean-field theory. Therefore, an important question regarding the modeling of unconventional superconductivity is asking where the electronic, magnetic, and lattice fluctuations appear in the formalism. 

Within the Green's function formalism, the mean-field solution is given by the Hartree-Fock approximation to the two-particle propagator, i.e., where the two-particle Green's function is approximated by the antisymmetric combination of single-particle Green's functions. In the case of the Gor'kov-Hedin-Baym equations, since the one-electron Green's function corresponds to an expansion in the screened potential rather than the bare Coulomb potential, the Hartree-Fock approximation corresponds to the GWA. So, contributions to $\mathbf{\Sigma}$ that go beyond the GWA constitute the fluctuations of the system.

\begin{figure}[t!]
\begin{center}
\includegraphics[width=0.90\columnwidth]{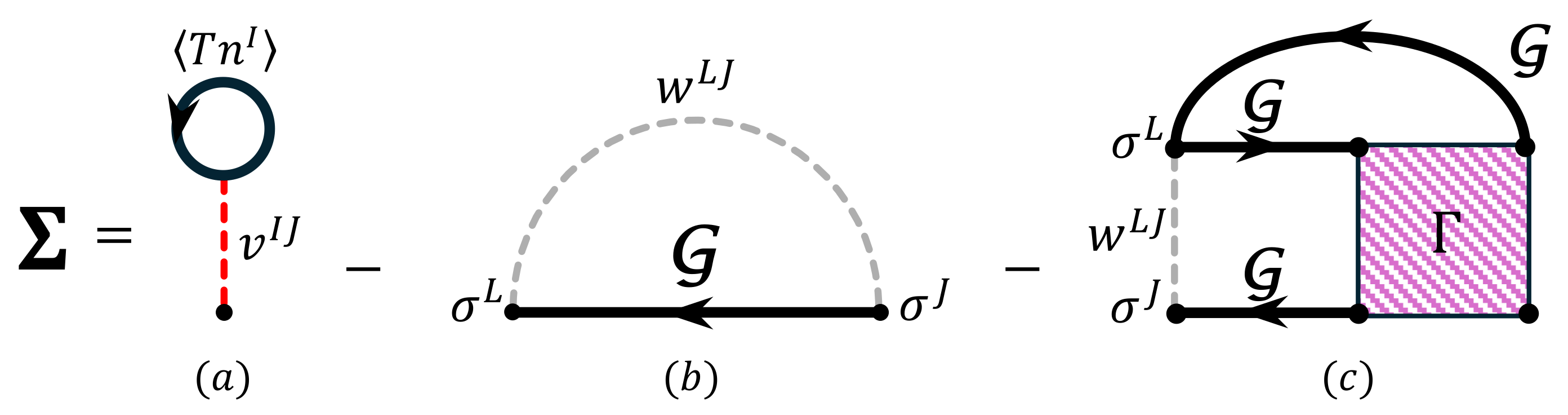}
\caption{  (color online) Diagrammatic representation of the exact self-energy with contributions from the (a) Hartree, (b) Fock/GW, and (c) effective interactions $\Gamma$ explicitly shown.} 
\label{fig:SelfEnergyEffectiveInt}
\end{center}
\end{figure}

Contributions that go beyond the GWA arise from interactions between quasiparticles and manifest as non-vanishing terms in $\frac{ \delta \mathbf{\Sigma} }{ \delta \pmb{\mathcal{G}}}$ of the two-particle vertex $\mathbf{\Lambda}$. These so-called vertex corrections account for exchange-correlation effects between an electron and the other electrons (holes) in the screening density cloud, which includes the electron-hole attraction in the dielectric response (exciton effects) and electron-electron attraction that gives way to the superconductivity. Such corrections become important for systems with strong electron-phonon coupling (breakdown of Migdal’s theorem)\cite{esterlis2018breakdown,mishra2025electron} or strong electronic and magnetic correlations\cite{nelson2007self,romaniello2009self,onida2002electronic}. 

To isolate the effective quasiparticle interactions, and thus the fluctuations, we re-sum $\mathbf{\Lambda}$ to group terms with $\frac{ \delta \mathbf{\Sigma} }{ \delta \pmb{\mathcal{G}}}$ yielding,
\begin{align}\label{eq:vertex2effectiveinteraction}
&\Lambda^{Lij}_{\mu\nu}(4,5;6)=\delta(4,6)\delta(4,5)\sigma^{Lij}_{\mu\nu}\nonumber\\
&+\Gamma^{ijab}_{\mu\nu\eta\xi}(5,6,11,12)
\mathcal{G}^{ac}_{\eta\tau}(11,6)
\sigma^{Lcd}_{\tau\epsilon}
\mathcal{G}^{db}_{\tau\xi}(6,12),
\end{align}
where $\Gamma$ is the effective interaction between quasiparticles:
\begin{align}
&\Gamma^{ijab}_{\mu\nu\eta\xi}(5,6,11,12)=\frac{\delta \Sigma^{ij}_{\mu\nu}(5,6) }{\delta \mathcal{G}^{ab}_{\eta\xi}(11,12) }\nonumber\\
&+
\frac{\delta \Sigma^{ij}_{\mu\nu}(5,6) }{\delta \mathcal{G}^{kl}_{\alpha\beta}(7,8) }
\mathcal{G}^{km}_{\alpha\gamma}(7,9)
\Gamma^{mnab}_{\gamma\delta\eta\xi}(9,10,11,12)
\mathcal{G}^{nl}_{\delta\beta}(10,8).
\end{align}
By substituting Eq.~\ref{eq:vertex2effectiveinteraction} into Eq.~\ref{eq:hedineqSE} and explicitly including the Hartree potential the exact self-energy is:
\begin{widetext}
\begin{align}\label{eq:SEwithfullvertex}
\Sigma^{sy}_{\eta\nu}(1,5)=\sigma^{Jsy}_{\eta\nu}
V_{H}^{J}(1)\delta(1,5)
-&
\sigma^{Jsi}_{\eta\gamma}
\mathcal{G}^{ix}_{\gamma\mu}(1,5)
\sigma^{Lxy}_{\mu\nu}
w^{LJ}(5,1)\nonumber\\
-&
\sigma^{Jsi}_{\eta\gamma}
\mathcal{G}^{ix}_{\gamma\mu}(1,4)
\Gamma^{xyab}_{\mu\nu\alpha\beta}(4,5,11,12)
\mathcal{G}^{ac}_{\beta\tau}(11,6)
\sigma^{Lcd}_{\tau\epsilon}
\mathcal{G}^{db}_{\epsilon\alpha}(6,12)
w^{LJ}(6,1)
\end{align}
\end{widetext}
The first two terms of Eq.~\ref{eq:SEwithfullvertex} are the Hartree and Fock components, respectively, where the Hartree potential couples the Nambu Green's function to the average total charge (spin) density of the electrons and nuclei in the system, and the Fock term is given by the GW (screened exchange interaction) self-energy. The third term represents all contributions beyond the independent particle Hartree-Fock approximation, where $\Gamma$ couples two quasiparticles and describes their multiple scattering processes that yield an effective (attractive or repulsive) potential. See Fig.~\ref{fig:SelfEnergyEffectiveInt} for a diagrammatic representation of Eq.~\ref{eq:SEwithfullvertex} .

\begin{figure}[t]
\begin{center}
\includegraphics[width=0.99\columnwidth]{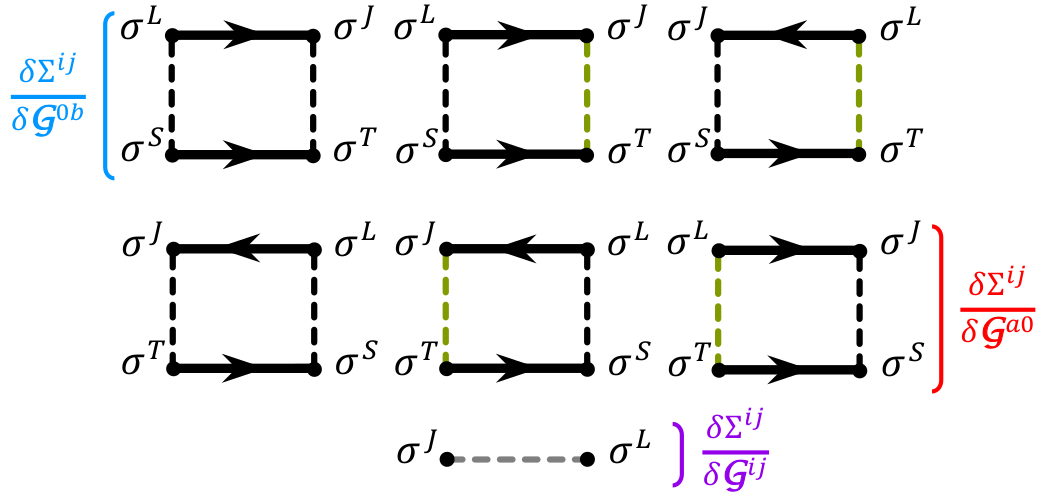}
\caption{(color online) Diagrammatic representation of the various contributions to the vertex in the GW approximation, where the solid black lines, dashed black lines, and green dashed lines represent the Gor'kov Green's function $(\pmb{\mathcal{G}})$, electronic screened interaction $(w_{e})$, and the phonon screened interaction $(w_{ph})$, respectively.   } 
\label{fig:Vertex} 
\end{center}
\end{figure}

The properties of $\frac{ \delta \mathbf{\Sigma}}{ \delta \pmb{\mathcal{G}}}$ and by extension $\Gamma$ encode information about potential Cooper pairs and the fluctuations at play\cite{schrieffer2018theory,maiti2013superconductivity}. As pointed out by Linscheid and Essenberger\cite{linscheid2015hedin} vertex contribution beyond the Hartree-Fock are needed to describe angular momentum transfer processes. That is, if we consider the special case of a pure Coulomb interaction $\left( w^{IJ}=w^{00}\delta_{IJ}\delta_{I0} \right)$ and spin-independent nuclei fluctuations $\left(D^{MN}=D^{00}\delta_{MN}\delta_{N0}\right)$, the Hartree and Fock self-energy do not capture spin-flip processes since $V^{0}_{H}$ and $w^{00}$ conserve angular momentum. Thus, these diagrams cannot explain the spin-fluctuation mediated pairing proposed for a broad class of unconventional superconducting materials\cite{scalapino2012common}. In general, $\Gamma$ can propagate angular momentum and other quantities that are conserved at the bare vertex. This is possible since a subset of diagrams that constitute the effective interaction have Green's function lines that do not connect across the interactions, thereby isolating loops of Green's functions. In the case of spin dependent interactions, $J=L+S$ angular momentum is conserved, thus yielding spin-flip processes of the total angular momentum. Furthermore, in the presence of a finite pairing field, these spin flips will also couple to Cooper pairs with finite angular momentum.

In general, determining $\Gamma$ is highly nontrivial and presents a grand challenging in condensed matter physics. Thus, many approximate schemes have been proposed that target or combine various fluctuation channels (subset of diagrams)\cite{essenberger2014density,maiti2013superconductivity,springer1998first,senechal2004theoretical,schindlmayr1998spectra,bickers1989conserving,del1994gwgamma,romaniello2012beyond,essenberger2014superconducting,onida2002electronic} to obtain an effective pairing potential. These approximations have seen various degrees of success in predicting the superconducting gap symmetry of various models and material systems\cite{scalapino1986d,scalapino2012common,kreisel2022superconducting,romer2019knight,hirschfeld2011gap,lane2022identifying,nandkishore2012chiral}.  

In the next section, a specific example of vertex corrections is presented based on the GW self-energy to illustrate various salient features. 

\subsubsection{Vertex Corrections}\label{sec:vertexcorrectionsLambda}

Using the GW approximation for the self-energy, we feed it back through the Gor'kov-Hedin-Baym equations following Fig.~\ref{fig:HEDINLOOP}(a). Now we obtain a non-vanishing functional derivative $\frac{\delta \mathbf{\Sigma}}{\delta \pmb{\mathcal{G}}}$ that yields corrections beyond the bare vertex. 
Inserting the GW self-energy into Eq.~\ref{eq:hedinVertex}, the vertex corrections are given by:
\begin{align}\label{eq:vertextypes}
\frac{ \delta \Sigma^{ij}_{\mu\nu}(5,6) }{ \delta  \mathcal{G}^{ab}_{\eta\xi}(11,12)  }=&
\frac{ \delta \Sigma^{ij}_{\mu\nu}(5,6) }{ \delta  \mathcal{G}^{0b}_{\eta\xi}(11,12)  }\delta_{a0}
+
\frac{ \delta \Sigma^{ij}_{\mu\nu}(5,6) }{ \delta  \mathcal{G}^{a0}_{\eta\xi}(11,12)  }\delta_{b0}\nonumber\\
&+
\frac{ \delta \Sigma^{ij}_{\mu\nu}(5,6) }{ \delta  \mathcal{G}^{ij}_{\eta\xi}(11,12)  }\delta_{ai}\delta_{bj}
\end{align}
where
\begin{widetext}
\begin{subequations}\label{eq:vertexcorrection}
\begin{align}
\frac{ \delta \Sigma^{ij}_{\mu\nu}(5,6) }{ \delta  \mathcal{G}^{0b}_{\eta\xi}(11,12)  }
=&
-w_{e}^{LS}(6,11)\sigma^{Sbq}_{\xi\beta}
\mathcal{G}^{qr}_{\beta\alpha}(12,11)\sigma^{Tr0}_{\alpha\eta}w_{e}^{TJ}(12,5)\sigma^{Jim}_{\mu\gamma}
\mathcal{G}^{mn}_{\gamma\epsilon}(5,6)\sigma^{Lnj}_{\epsilon\nu}\nonumber\\
&-w_{e}^{LS}(6,11)\sigma^{Sbq}_{\xi\beta}
\mathcal{G}^{qr}_{\beta\alpha}(12,11)\sigma^{Tr0}_{\alpha\eta}w_{ph}^{TJ}(12,5)\sigma^{Jim}_{\mu\gamma}
\mathcal{G}^{mn}_{\gamma\epsilon}(5,6)\sigma^{Lnj}_{\epsilon\nu}\nonumber\\
&-w_{e}^{JS}(5,11)\sigma^{Sbq}_{\xi\beta}
\mathcal{G}^{qr}_{\beta\alpha}(12,11)\sigma^{Tr0}_{\alpha\eta}w_{ph}^{LT}(6,12)\sigma^{Jim}_{\mu\gamma}
\mathcal{G}^{mn}_{\gamma\epsilon}(5,6)\sigma^{Lnj}_{\epsilon\nu},
\\
\frac{ \delta \Sigma^{ij}_{\mu\nu}(5,6) }{ \delta  \mathcal{G}^{a0}_{\eta\xi}(11,12)  }
=&
-w_{e}^{LS}(6,12)\mathcal{G}^{0p}_{\delta\alpha}(12,11)
\sigma^{Spa}_{\alpha\eta}\sigma^{T00}_{\xi\delta}w_{e}^{TJ}(11,5)\sigma^{Jim}_{\mu\gamma}
\mathcal{G}^{mn}_{\gamma\epsilon}(5,6)\sigma^{Lnj}_{\epsilon\nu}\nonumber\\
&-w_{e}^{LS}(6,12)\mathcal{G}^{0p}_{\delta\alpha}(12,11)
\sigma^{Spa}_{\alpha\eta}\sigma^{T00}_{\xi\delta}w_{ph}^{TJ}(11,5)\sigma^{Jim}_{\mu\gamma}
\mathcal{G}^{mn}_{\gamma\epsilon}(5,6)\sigma^{Lnj}_{\epsilon\nu}\nonumber\\
&-w_{e}^{JS}(5,12)\mathcal{G}^{0p}_{\delta\alpha}(12,11)
\sigma^{Spa}_{\alpha\eta}\sigma^{T00}_{\xi\delta}w_{ph}^{LT}(6,11)\sigma^{Jim}_{\mu\gamma}
\mathcal{G}^{mn}_{\gamma\epsilon}(5,6)\sigma^{Lnj}_{\epsilon\nu},
\\
\frac{ \delta \Sigma^{ij}_{\mu\nu}(5,6) }{ \delta  \mathcal{G}^{ij}_{\eta\xi}(11,12)}
=&-w^{LJ}(5,6)   \sigma^{Jii}_{\mu\eta} \sigma^{Ljj}_{\xi\nu} \delta(5,11)\delta(6,12). 
\end{align}
\end{subequations}
\end{widetext}
The Nambu indices have been written explicitly for clarity. Figure~\ref{fig:Vertex} presents a diagrammatic representation of the various terms in Eq.~\ref{eq:vertexcorrection}. Three distinct terms are produced: (a) two-rung particle-particle ladders, (b) two-rung particle-hole ladders, and (c) an exchange interaction. In general, the two-particle effective interaction $\frac{ \delta \mathbf{\Sigma}_{\mu\nu} }{ \delta \pmb{\mathcal{G}}_{\alpha\beta}}$ has 16 Nambu components that couple the various ordinary and anomalous sectors, along with the charge, spin, and lattice degrees of freedom. Here, only 13 terms are nonzero originating from the polarizability being scalar in Nambu space and therefore forcing vertices to conserve the number of incoming and outgoing lines. 

If we were to continue to iterate the Gor'kov-Hedin-Baym equations, the number of two-point vertex graphs grows factorially. In particular, the variety of exchange and direct graphs would emerge that form the basis for the T-matrix approximation\cite{keller1999thermodynamics}, Kohn-Luttinger\cite{luttinger1966new,kohn1965new}, FLEX\cite{bickers1989conservingI,esirgen1997fluctuation,bickers1989conserving}, RPA\cite{romer2015pairing,scalapino1986d,lane2022identifying} approximations, and  the parqet equations\cite{eckhardt2023functional,bickers2004self}. Furthermore, various local, cluster, non-local diagrammatic approximations (e.g., DMFT\cite{georges1996dynamical,kotliar2006electronic}, CDMFT\cite{park2008cluster,maier2005quantum,tremblay2006pseudogap,senechal2011cluster}, DMFT+GW\cite{sun2002extended,biermann2003first}, DCA\cite{hettler1998nonlocal,maier2005quantum}, VCA\cite{potthoff2003variational}, D$\Gamma$A\cite{toschi2007dynamical,held2008dynamical}) to the self-energy and vertex are able to capture short- to long-range fluctuations. The T-matrix and RPA approximations can be constructed within the Gor'kov-Hedin-Baym framework by following Ref.~\onlinecite{romaniello2012beyond} and \onlinecite{lane2022identifying}. Finally, we wish to point out that by iterating the Gor'kov-Hedin equations effective interactions arise between anomalous Green's functions in addition to the particle-particle and particle-hole interactions between ordinary propagators. That is, despite the bare interactions conserving particle number, higher-order effective interactions facilitate virtual particle non-conserving processes.

\begin{figure}[t]
\begin{center}
\includegraphics[width=0.90\columnwidth]{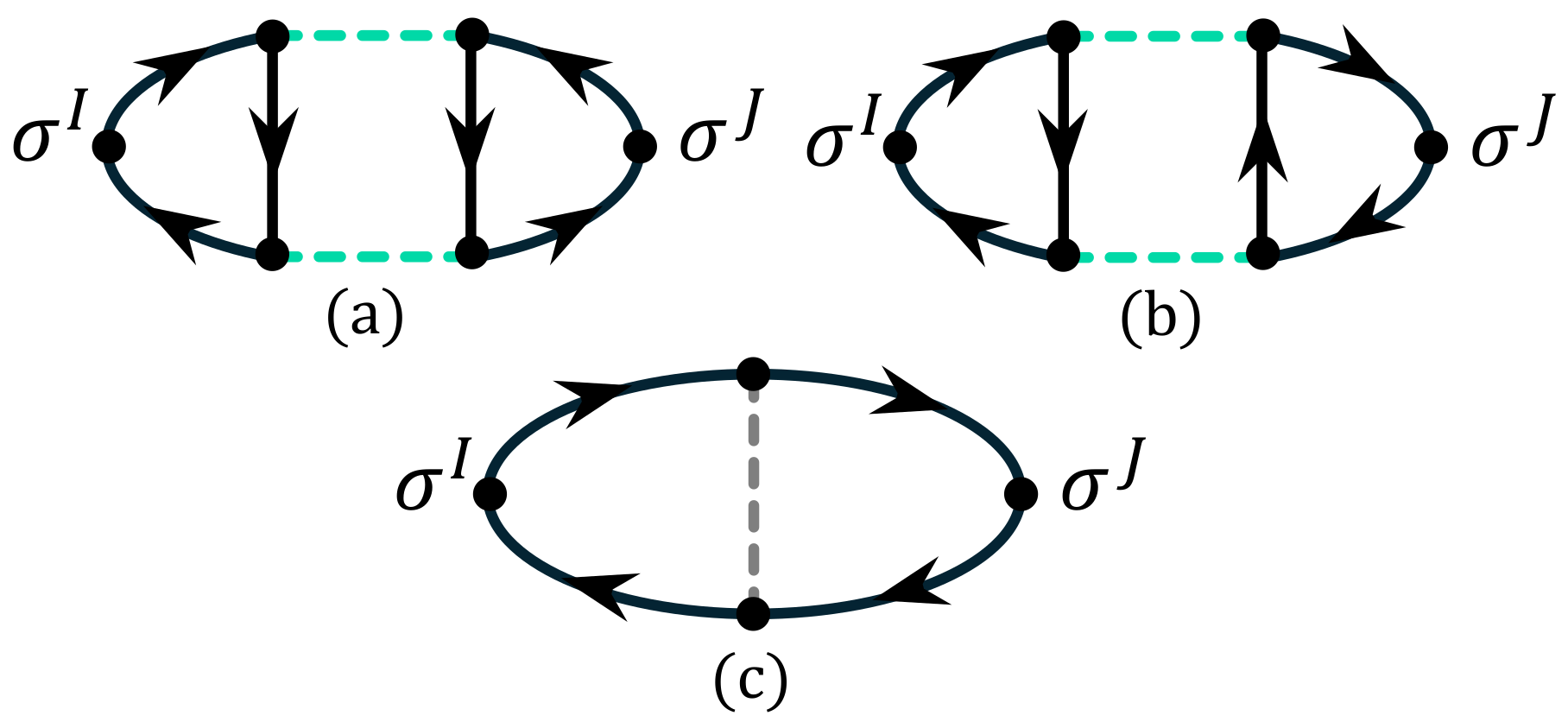}
\caption{ (color online) Diagrammatic representation of the various contributions to the vertex corrected polarizability: (a) particle-particle and (b) particle-hole Aslamazov-Larkin type graphs, and (c) a Maki-Thompson type diagram, where the solid black lines and dashed gray lines represent the Gor'kov Green's function $(\pmb{\mathcal{G}})$ and screened interaction $(w)$, respectively. The dashed teal lines represent either the electronic screened interactions $(w_{e})$ or the screened phonon contribution $(w_{ph})$, where only one screened phonon interaction can be inserted into a polarizability bubble at a time.} 
\label{fig:PolizationVertex} 
\end{center}
\end{figure}

\subsubsection{Vertex Corrected Polarizability and Screened Interaction}

To complete the second iteration of the Gor'kov-Hedin-Baym equations, the updated vertex $\mathbf{\Lambda}$ is inserted into $p_{e}$. Now, the polarizability,
\begin{align}
p^{MN}_{e}(7,8)=&
\mathcal{G}^{0i}_{\delta\mu}(7,8)
\sigma^{Nij}_{\mu\nu}
\mathcal{G}^{jm}_{\nu\eta}(8,7^+)
\sigma^{Mm0}_{\eta\delta}\nonumber\\
+&
\mathcal{G}^{0i}_{\delta\mu}(7,9)
\frac{ \delta \Sigma^{ij}_{\mu\nu}(9,10) }{ \delta  \mathcal{G}^{ab}_{\alpha\beta}(11,12)}
\mathcal{G}^{jl}_{\nu\eta}(10,7^+)
\sigma^{Ml0}_{\eta\delta}\nonumber\\
&\times\mathcal{G}^{am}_{\alpha\sigma}(11,8)
\sigma^{Nmn}_{\sigma\epsilon}
\mathcal{G}^{nb}_{\epsilon\beta}(8,12),
\end{align}
is composed of two parts: a non-interacting bubble that was present in the RPA and new terms where the Nambu Green's functions are coupled via the effective interactions given in Eq.~\ref{eq:vertextypes}. Specifically, the interacting bubbles take the form of the well known Aslamazov-Larkin\cite{aslamazov1968effect,aslamasov1968influence,larkin2005theory} and Maki-Thompson\cite{maki1968critical,thompson1970microwave} contributions to the polarizability, depicted in Fig.~\ref{fig:PolizationVertex}. Importantly these contributions provide a key to going beyond the BCS mean-field theory of superconductivity by laying the foundation of the microscopic theory of superconducting fluctuations in the normal phase of a superconductor. The generalized graphs presented here extend this theory to spin-dependent electronic and phonon interactions. These contributions to the polarizability, and more generally the electronic response function, are crucial to modeling dynamic response of spectroscopes via inelastic light scatting\cite{devereaux2007inelastic}, transport\cite{aslamasov1980fluctuation}, ultrasound attenuation\cite{mar2004superconducting}, to name a few. Furthermore, these contributions are a generalization of those employed by Anderson\cite{anderson1958random}, Bogoliubov, Tolmachev, and Shirkov\cite{bogoliubov1959new}, Vaks, Galitskii, and Larkin\cite{vaks1962collective}, and Bardasis and Schrieffer\cite{bardasis1961excitons} to investigate collective excitations (e.g. plasmons, excitons) and life-time effects in BCS superconductors. Such effects of fluctuations directly influence the screed interaction and therefore, plays a key role in modifying the dynamical phonon self-energy and electron-phonon coupling, in addition to the electron self-energy. 

\subsubsection{The Self Energy}

\begin{figure}[t]
\begin{center}
\includegraphics[width=0.99\columnwidth]{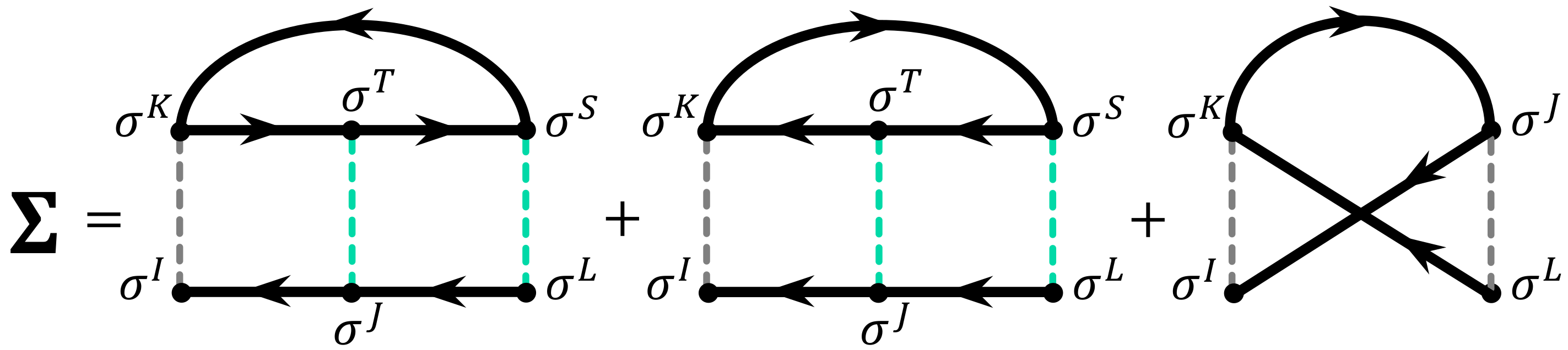}
\caption{(color online) Diagrammatic representation of the various contributions to the self-energy including vertex corrections within the GW approximation, where the solid black lines, dashed black lines, and dashed gray lines, represent the Gor'kov Green's function $(\pmb{\mathcal{G}})$, the electronic screened interaction $(w_{e})$, and the total screened interaction $(w)$. The dashed teal lines represent either the electronic screened interactions $(w_{e})$ or the screened phonon contribution $(w_{ph})$, where only one screened phonon interaction can be inserted at a time.} 
\label{fig:SEVertex} 
\end{center}
\end{figure}

By iterating beyond the GW approximation the various self-energy diagrams now include three-rung particle-hole ladders, three-rung particle-particle ladders, and the screened second-order exchange graph. Figure~\ref{fig:SEVertex} presets the diagrammatic representation of the self-energy including vertex corrects. Due to the matrix form the Nambu Green's function and the effective interactions, the three-rung ladders describe three scenarios: (i) a particle or hole in the many-body system scatters multiple times off of another particle $G~(\bar{G})$ in the system, (ii) a particle or hole scatters multiple times off of the condensate $F~(\bar{F})$ in the system, and (iii) the condensate interacts multiple times with itself $F~(\bar{F})$. Moreover, these interactions are mediated by the exchange of plasmons $(w_{e}^{00})$, magnons $(w_{e}^{ij})$, and phonons $(w_{ph}^{IJ})$, thus the spin structure of the condensate is highly dependent on the existence of the relativistic effects, similar to the GW case. 

If we assume the Green's function is diagonal in spin space and the pairing field is zero, the three-rung ladder self-energies capture spin and charge fluctuations as discussed by Doniach and Engelsberg\cite{doniach1966low} and Larkin and Varlamov\cite{larkin2005theory}. In the presence of strong spin-orbit coupling, the present formulation naturally generalizes this scenario to permit the propagation of angular momentum $J=L+S$, or in other words, spin-flip processes of the total angular momentum. For finite pairing fields, condensates with a angular momentum greater than zero, e.g. $J=1$, $2$, $3$, etc, can also mediate spin-flip processes.

\section{Connecting to First-Principles Calculations}
To connect the various electron-phonon coupling terms to first-principles calculations, e.g., density functional theory calculations, we express the displacements $\Delta\mathbf{\tau}^{i}_{\kappa p}$ in the normal vibration modes of the crystal
\begin{align}
\Delta \mathbf{\tau}^{i}_{\kappa p}(\tau)&=
\left(
\frac{\hbar}{2N_{p}M_{\kappa}\omega_{\mathbf{q}\nu}}\right)^{1/2}\nonumber\\
&\times\sum_{\mathbf{q}\nu} e^{i\mathbf{q}\cdot \mathbf{R}_{p}} \hat{\varepsilon}^{i}_{\kappa \nu}(\mathbf{q}) \left[ a_{\mathbf{q}\nu}(\tau)+a^{\dagger}_{-\mathbf{q}\nu}(\tau) \right]
\end{align}
where $N_{p}$ is the number of unit cells, $M_{\kappa}$ is the mass of the $\kappa$ basis atom, $\hat{\varepsilon}^{i}_{\kappa \nu}(\mathbf{q})$ is the polarization of mode $\nu$ of momentum $\mathbf{q}$, and $a_{\mathbf{q}\nu}\left(a^{\dagger}_{\mathbf{q}\nu}\right)$ is the annihilation (creation) operator for each phonon of energy $\omega_{\mathbf{q}\nu}$. The phonon Green's function is now given by
\begin{align}
\mathcal{D}_{\mathbf{q}\nu}(7,10) &= -\braket{A_{\mathbf{q}\nu}(7)A_{-\mathbf{q}\nu}(10)},
\end{align}
and the first-order electron-phonon coupling matrices
\begin{align}
g^{M}_{\mathbf{q}\nu}(8,7) &=\sum_{\kappa p \alpha}
\left( \frac{\hbar}{2N_{p}M_{\kappa}\omega_{\mathbf{q}\nu}}  \right)^{1/2}  \varepsilon^{-1~ML}_{e}(8,6)\nonumber\\
&\times e^{i\mathbf{q}\cdot \mathbf{R}_{p}} \hat{\varepsilon}^{i}_{\kappa \nu}(\mathbf{q}) X^{L}_{\kappa} \nabla^{i}_{6} v(6-\mathbf{\tau}^{0}_{\kappa p}),\\
g^{J}_{-\mathbf{q}\nu}(1,10) &=\sum_{\kappa^{\prime} p^{\prime} \alpha^{\prime}}
\left( \frac{\hbar}{2N_{p^{\prime}}M_{\kappa^{\prime}}\omega_{\mathbf{-q}\nu}}  \right)^{1/2}  \varepsilon^{-1~JA}_{e}(1,5) \nonumber\\
&\times e^{-i\mathbf{q}\cdot \mathbf{R}_{p^{\prime}}} \hat{\varepsilon}^{i^{\prime}}_{\kappa^{\prime} \nu}(-\mathbf{q}) X^{A}_{\kappa^\prime}\nabla^{i^{\prime}}_{5} v(5-\mathbf{\tau}^{0}_{\kappa^{\prime} p^{\prime}}),
\end{align}
with $w_{ph}$ now given by
\begin{align}
w_{ph}^{MJ}(8,1) &= 
\sum_{\mathbf{q}\nu}
g^{M}_{\mathbf{q}\nu}(8,7) 
\bar{D}_{\mathbf{q}\nu}(7,10)
g^{J}_{-\mathbf{q}\nu}(1,10).
\end{align}
The second-order electron-phonon coupling matrices are now
\begin{align}
g^{(2)J}_{\mathbf{q}\nu\mathbf{q}^\prime\nu^\prime}(1)=
\sum_{\kappa p}&
\left( \frac{\hbar}{4N_{p}M_{\kappa}\sqrt{\omega_{\mathbf{q}\nu}\omega_{\mathbf{q}^\prime\nu^\prime}}}  \right) \nonumber\\
&\times X^I_{\kappa p}\nabla^i_{\mathbf{r}_2}\nabla^j_{\mathbf{r}_2}v^{IJ}(1-\pmb{\tau}^0_{\kappa p})
\end{align}
and the Debye-Waller self-energy can be written as
\begin{align}
V_{DW}^{J}(1)&=-
\sum_{\mathbf{q}\mathbf{q}^\prime\nu\nu^\prime}\mathcal{D}_{\mathbf{q}\nu\mathbf{q}^\prime\nu^\prime}(\tau_2,\tau_2)g^{(2)J}_{\mathbf{q}\nu\mathbf{q}^\prime\nu^\prime}(1)
\end{align}
Comparing to Marini {\it et al.}\cite{marini2015many}, $g^{M}$ and $g^{(2)J}$ are equivalent to $\xi$ and $\theta$ in Ref.~\onlinecite{marini2015many}, with $\Xi$ in Ref.~\onlinecite{marini2015many} equivalent to the first term in Eq.~\ref{eq:SEphonon}.
For more details on the phonon Green's function and electron-phonon coupling matrices, see Refs.~\onlinecite{marini2015many,baym1961field,bernardi2016first,giustino2017electron,hedin1970effects,maksimov1975self}.

\section{Conclusion}
In conclusion, we have extended the original set of Hedin equations for many-electron systems with purely Coulombic interactions in a vibrating lattice to systems with explicitly spin-dependent interactions and finite pairing fields. This framework provides a natural platform to examine the interplay of correlations and electron-lattice coupling in the presents of strong relativistic effects in materials specific detail. Such an approach helps to address challenges in accurately determining the presence of topological superconductivity in known candidate materials and find guiding principles for materials discovery efforts.

\section*{Acknowledgements}
This work was carried out under the auspices of the US Department of Energy (DOE) National Nuclear Security Administration under Contract No. 89233218CNA000001. It was supported by the Quantum Science Center, a U.S. DOE Office of Science National Quantum Information Science Research Center.

\begin{appendix}
\numberwithin{equation}{section}

\section{Derivation of the Gor'kov-Hedin Equations}\label{Appendix:derivationGH}
The derivation of the Gor'kov-Hedin equations closely follows Hedin's original work using Schwinger's functional derivative technique. Using the Heisenberg equation of motion 
\[
\frac{d}{d\tau_{1}}\hat{\psi}_{\eta}(1)=\left[  \mathcal{H}, \hat{\psi}_{\eta}(1) \right] \qquad\mbox{and}\qquad
\frac{d}{d\tau_{1}}\hat{\psi}^{\dagger}_{\eta}(1)=\left[  \mathcal{H}, \hat{\psi}^{\dagger}_{\eta}(1) \right],
\]
for annihilation and creation operators, respectively, we obtain after computing the commutator
\begin{widetext}
\begin{align}
\frac{d}{d\tau_{1}}\hat{\psi}_{\eta}(1)=&
-\sum_{\beta}h_{\eta\beta}(1)\hat{\psi}_{\beta}(1)
-\sum_{I\beta}\pi^I(1)\sigma^{I}_{\eta\beta}(1)\hat{\psi}_{\beta}(1) 
-\sum_{IJ\gamma}\int \mathrm{d}3n^{I}(3)v^{IJ}(3,1)\sigma^{J}_{\eta\gamma}\hat{\psi}_{\gamma}(1)
+\sum_{\alpha}\int \mathrm{d}3\bar{\Delta}_{\eta\alpha}(1,3)\hat{\psi}^{\dagger}_{\alpha}(3),
\\
\frac{d}{d\tau_{1}}\hat{\psi}^{\dagger}_{\eta}(1)=&
\sum_{\alpha}h^{*}_{\eta\alpha}(1)\hat{\psi}^{\dagger}_{\alpha}(1)
+\sum_{I\alpha}\pi^{I}(1)\sigma^{*I}_{\eta\alpha}\hat{\psi}^{\dagger}_{\alpha}(1) 
+\sum_{IJ\gamma}\int \mathrm{d}3 v^{IJ}(3,1)\sigma^{*J}_{\eta\gamma}\hat{\psi}^{\dagger}_{\gamma}(1)n^{I}(3)
-\sum_{\alpha}\int \mathrm{d}3\Delta_{\eta\alpha}(1,3)\hat{\psi}_{\alpha}(3),
\end{align}
\end{widetext}
where $\pi^I(1)$ is an auxiliary field and will be taken to zero at the end of the analysis. Multiplying by an annihilation (creation) operator from the right and taking the time ordered ensemble average, we may write the resulting equation of motion in terms of ordinary and anomalous Green's functions defined as:
\begin{subequations}
\begin{align}
G_{\eta\xi}(1,2)=-\braket{  \mathcal{T} \left\{ \hat{\psi}_{\eta}(1) \hat{\psi}^{\dagger}_{\xi}(2) \right\}  }\\
F_{\eta\xi}(1,2)=-\braket{  \mathcal{T} \left\{ \hat{\psi}_{\eta}(1) \hat{\psi}_{\xi}(2) \right\}  }\\
\bar{F}_{\eta\xi}(1,2)=-\braket{  \mathcal{T} \left\{ \hat{\psi}^{\dagger}_{\eta}(1) \hat{\psi}^{\dagger}_{\xi}(2) \right\}  }\\
\bar{G}_{\eta\xi}(1,2)=-\braket{  \mathcal{T} \left\{ \hat{\psi}^{\dagger}_{\eta}(1) \hat{\psi}_{\xi}(2) \right\}  }
\end{align}
\end{subequations}
which are the components of the 
matrix propagator 
\begin{align}
\mathfrak{G}_{\eta\xi}(1,2)=-
\left\langle{\mathcal{T}\Psi_{\eta}(1)\otimes\Psi^{\dagger}_{\xi}(2)}\right\rangle.
\end{align}
For the ease of notation we define 
\begin{align}
\pmb{\mathcal{G}}_{\eta\xi}(1,2) =&\tau^{z}\mathfrak{G}_{\eta\xi}(1,2)\nonumber\\
=&\left[\begin{array}{cc}
G_{\eta\xi}(1,2) &F_{\eta\xi}(1,2)\\
-\bar{F}_{\eta\xi}(1,2)& -\bar{G}_{\eta\xi}(1,2) 
\end{array}\right],\nonumber
\end{align}
where $\tau^{z}$ is equivalent to $\sigma^{z}$ except it acts on the Nambu components of $\mathfrak{G}$. With these definitions, we arrive to the equation of motion of the matrix Nambu Green’s function:
\begin{align}\label{eq:eom_full}
\pmb{\mathcal{G}}^{-1}_{0\eta\beta}(1,3)\pmb{\mathcal{G}}_{\beta\xi}(3,2)
=&\pmb{\sigma}^{0}_{\eta\xi}\delta(1,2)\nonumber\\
&+v^{IJ}(3,1)\pmb{\sigma}^{J}_{\eta\gamma}\pmb{\mathcal{G}}^{(2)I}_{\gamma\xi}(1,3,2),
\end{align}
with 
\begin{align}
&\pmb{\mathcal{G}}^{(2)I}_{\gamma\xi}(1,3,2)=\nonumber\\
&\begin{bmatrix}
-\left\langle{\mathcal{T}\left\{ n^{I}(3)\hat{\psi}_{\gamma}(1) \hat{\psi}^{\dagger}_{\xi}(2)  \right\}   }\right\rangle
&
-\left\langle{\mathcal{T}\left\{ n^{I}(3)\hat{\psi}_{\gamma}(1) \hat{\psi}_{\xi}(2)  \right\}   }\right\rangle
\\
\left\langle{\mathcal{T}\left\{ \hat{\psi}^{\dagger}_{\gamma}(1)n^{I}(3) \hat{\psi}^{\dagger}_{\xi}(2)  \right\}   }\right\rangle
&
\left\langle{\mathcal{T}\left\{ \hat{\psi}^{\dagger}_{\gamma}(1)n^{I}(3) \hat{\psi}_{\xi}(2)  \right\}   }\right\rangle
\end{bmatrix},
\end{align}
and the block Pauli matrix $\pmb{\sigma}^{J}_{\eta\gamma}$ defined as 
\begin{align}\label{eq:blockpauli}
\left[\begin{array}{cc}
\sigma^{J}_{\eta\gamma} & 0\\
0 & \sigma^{*J}_{\eta\gamma}
\end{array}\right].
\end{align}

To utilize the Schwinger functional derivative technique, 
we define all operators within the {\it imaginary-time} Heisenberg picture,
\begin{align}
\mathcal{O}(z)=U(\tau_0,\tau)\mathcal{O}U(\tau,\tau_0),
\end{align}
where the time arguments $\tau,\tau_0$ run along the imaginary-axis of the Keldysh contour. The time-evolution operator $U(\tau,\tau_0)$ evolves a given operator $\mathcal{O}$ from an arbitrary initial time $\tau_{0}$ to $\tau$ along the imaginary-axis. Here, the operators are explicitly time dependent, unlike the Schr\"odinger picture where the wave functions are time dependent. To treat the electronic many-body dynamics at finite temperature, we may define the time-dependent ensemble average of operator $\mathcal{O}(\tau)$ as 
\begin{align}
\left\langle{\mathcal{O}(\tau)}\right\rangle=\frac{\mbox{Tr}\left\{  \mathcal{T} \exp{\left[-\int_{0}^{\beta}  d\bar{\tau} H(\bar{\tau}) \right]    }    \mathcal{O}(\tau)  \right\}    }{\mbox{Tr}\left\{  \mathcal{T} \exp{\left[-\int_{0}^{\beta}  d\bar{\tau} H(\bar{\tau}) \right]    }    \right\}   },
\end{align}
where $\mathcal{T}$ is the imaginary-time-ordering operator, and $\left\langle{\mathcal{O}(\tau)}\right\rangle$ is the overlap between the initial state in thermodynamical equilibrium (for temperature $\beta$) at $\tau_0$ with the time evolved state at $\tau$. In Hedin's original work~\cite{hedin1965} he utilized a perturbing auxiliary electric potential to relate the two-particle and single-particle Green's function via the functional derivative. This auxiliary potential serves as a mathematical trick and is set to zero at the end of the calculation to yield the set of self consistent equations. Here, to capture both electronic and magnetic responses of the system we consider an auxiliary electric potential $\pi^{0}\equiv \phi_0$ and auxiliary magnetic field $\pi^{i}\equiv \frac{1}{2}g\mu_B B^i$, for $i=x,y,z$, where $\mathbf{B}=\nabla \times \mathbf{A}$ and $\mathbf{A}$ is the vector potential. The coupling between these auxiliary fields and our system is given as 
\begin{align}\label{eq:extfield}
\hat{\pi}(\tau)=&
\int \mathrm{d}\mathbf{r} \left[n(\mathbf{r},\tau) \phi_0(\mathbf{r},\tau)+g\mu_B \mathbf{B}(\mathbf{r},\tau)\cdot\mathbf{S}(\mathbf{r},\tau) \right]\nonumber\\
\equiv&\sum_I\int \mathrm{d}\mathbf{r} ~ \pi^{I}(\mathbf{r},\tau) n^{I}(\mathbf{r},\tau),
\end{align}
where $n$ is the total charge and $\mathbf{S}$ is the total spin magnetic moment.
It then can be shown~\cite{kadanoff_quantum_1962} that the change in the ensemble average of a generic, imaginary-time-ordered product of operators $\Pi_{i}\mathcal{O}_{i}(\tau_{i})$ with respect to field $\pi^{I}$ along the imaginary-time-axis yields
\begin{align}\label{eq:funcderivindentity}
-\frac{\delta}{\delta \pi^{I}(1)}\left\langle{\mathcal{T}\left\{  \Pi_{i}\mathcal{O}_{i}(\tau_{i})  \right\}   }\right\rangle=
\left\langle{\mathcal{T}\left\{  \Pi_{i}\mathcal{O}_{i}(\tau_{i}) n^{I}(1)  \right\}   }\right\rangle \nonumber\\
-\left\langle{\mathcal{T}\left\{  \Pi_{i}\mathcal{O}_{i}(\tau_{i})  \right\}   }\right\rangle \left\langle{\mathcal{T}\left\{     n^{I}(1)   \right\}   }\right\rangle.
\end{align}
From this expression it is clear that if $\mathcal{O}_{i}(\tau_{i})$ is a quadratic combination of creation (annihilation) operators, the right hand side is the difference between a two-particle Green's function and the multiplication  of two single-particle Green's functions. This inportant relation enables us to rewrite $\pmb{\mathcal{G}}^{(2)}$ in terms of $\pmb{\mathcal{G}}$ and its functional derivatives:
\begin{align}\label{eq:trick}
\pmb{\mathcal{G}}^{(2)I}_{\gamma\xi}(1,3,2)
=
\pmb{\mathcal{G}}_{\gamma\xi}(1,2)
\left\langle{\mathcal{T}\left\{     n^{I}(3)   \right\}   }\right\rangle
-\frac{\delta \pmb{\mathcal{G}}_{\gamma\xi}(1,2)}{\delta \pi^{I}(3)}.
\end{align}
Using Eq.~\ref{eq:trick} we can define the mass operator $\pmb{\mathcal{M}}$ as 
\begin{widetext}
\begin{align}
\pmb{\mathcal{M}}_{\eta\nu}(1,5)\pmb{\mathcal{G}}_{\nu\xi}(5,2)=&
v^{IJ}(3,1)
\pmb{\sigma}^{J}_{\eta\gamma}
\pmb{\mathcal{G}}^{(2)I}_{\gamma\xi}(1,3,2)\\
&=v^{IJ}(3,1)\pmb{\sigma}^{J}_{\eta\gamma}
\left[\pmb{\mathcal{G}}_{\gamma\xi}(1,2)\left\langle{\mathcal{T}\left\{     n^{I}(3)   \right\}   }\right\rangle-\frac{\delta \pmb{\mathcal{G}}_{\gamma\xi}(1,2)}{\delta \pi^{I}(3)} \right]\nonumber\\
&=v^{IJ}(3,1)\pmb{\sigma}^{J}_{\eta\gamma}
\left[\pmb{\mathcal{G}}_{\gamma\xi}(1,2) \left\langle{\mathcal{T}\left\{     n^{I}(3)   \right\}   }\right\rangle+
\pmb{\mathcal{G}}_{\gamma\mu}(1,4)
\frac{\delta   \pmb{\mathcal{G}}^{-1}_{\mu\nu}(4,5)  }{\delta \pi^{I}(3)} 
\pmb{\mathcal{G}}_{\nu\xi}(5,2)
\right]\nonumber\\
&=\left[V^{J}_{H}(1)\pmb{\sigma}^{J}_{\eta\gamma}\delta_{\gamma\nu}\delta(1,5)
+
v^{IJ}(3,1)\pmb{\sigma}^{J}_{\eta\gamma}\pmb{\mathcal{G}}_{\gamma\mu}(1,4)
\frac{\delta   \pmb{\mathcal{G}}^{-1}_{\mu\nu}(4,5)  }{\delta \pi^{I}(3)} 
\right] \pmb{\mathcal{G}}_{\nu\xi}(5,2). \nonumber
\end{align}
\end{widetext}
We recognize the first term of $\pmb{\mathcal{M}}$ as the Hartree potential and the second term as the exact expression for the self-energy:
\begin{align}\label{eq:exactSE}
\mathbf{\Sigma}_{\eta\nu}(1,5)&=
v^{IJ}(3,1)\pmb{\sigma}^{J}_{\eta\gamma}\pmb{\mathcal{G}}_{\gamma\mu}(1,4)
\frac{\delta   \pmb{\mathcal{G}}^{-1}_{\mu\nu}(4,5)  }{\delta \pi^{I}(3)}.
\end{align}
One of the goals of Hedin's original work was to derive a set of successively self-consistent equations for the one-electron Green's function that correspond to an expansion in the screened potential rather than the bare Coulomb potential. To achieve this, one takes the functional derivatives of $\pmb{\mathcal{G}}^{-1}$ with respect to the total field $\Phi^{I}=V_{H}^{I}+\pi^{I}$ instead of the bare perturbing potential $\pi^{I}$ via the chain rule:
\begin{align}\label{eq:vertexscreened}
\frac{\delta   \pmb{\mathcal{G}}^{-1}_{\mu\nu}(4,5)  }{\delta \pi^{I}(3)} =
\frac{\delta   \pmb{\mathcal{G}}^{-1}_{\mu\nu}(4,5)  }{\delta \Phi^{L}(6)}
\frac{\delta   \Phi^{L}(6)  }{\delta \pi^{I}(3)}.
\end{align}
The vertex function can be defined as 
\begin{align}\label{eq:vertexdefinition}
\mathbf{\Lambda}^{L}_{\mu\nu}(4,5;6)=-\frac{\delta   \pmb{\mathcal{G}}^{-1}_{\mu\nu}(4,5)  }{\delta \Phi^{I}(6)},
\end{align}
and the dielectric function is the derivative of the total field with respect to the applied one
\begin{align}
\varepsilon^{-1}_{LI}(6,3)&=\frac{\delta   \Phi^{L}(6)  }{\delta \pi^{I}(3)}=\delta(6,3)\delta_{IL}+\frac{\delta V^{L}_{H}(6) }{\delta \pi^{I}(3)}
\end{align}
Inserting Eq.~\ref{eq:vertexscreened} into Eq.~\ref{eq:exactSE} we obtain Eq.~\ref{eq:hedineqSE} with $w^{LJ}(6,1) = \varepsilon^{-1}_{LI}(6,3)v^{IJ}(3,1)$. The vertex equations [Eq.~\ref{eq:hedinVertex}] are given by
\begin{align}
&\mathbf{\Lambda}^{L}_{\mu\nu}(4,5;6)=-\frac{\delta \pmb{\mathcal{G}}^{-1}_{H~\mu\nu}(4,5) }{\delta \Phi^{L}(6)}+\frac{\delta \mathbf{\Sigma}_{\mu\nu}(4,5) }{\delta \Phi^{L}(6)}\nonumber\\
&=\delta(6,4)\delta(4,5)\pmb{\sigma}^{L}_{\mu\nu}\nonumber\\
&+\frac{\delta \mathbf{\Sigma}_{\mu\nu}(4,5) }{\delta \mathcal{G}^{ij}_{\alpha\beta}(9,10) }
\mathcal{G}^{im}_{\alpha\gamma}(9,11)
\Lambda^{L~mn}_{\gamma\eta}(11,12;6)
\mathcal{G}^{nj}_{\eta\beta}(12,10)
\end{align}
where we have made use of the chain rule and the definition of the vertex in Eq.~\ref{eq:vertexdefinition}. 

To establish the expression for the screened interaction, we start by identifying the electronic polarization as the variation of the density with respect to the total potential:
\begin{align}
&p^{KM}_{e}(7,8)=\frac{\delta \left\langle{\mathcal{T}\left\{     n^{K}_{e}(7)   \right\}   }\right\rangle}{\delta \Phi^{M}(8) }\nonumber\\
&=\frac{\delta {G}_{\delta\alpha}(7,7^{+})  }{\delta \Phi^{M}(8) } \sigma^{K}_{\alpha\delta}\nonumber\\
&=\left[ \frac{ \delta \pmb{\mathcal{G}}_{\delta\alpha}(7,7^{+})  }{\delta \Phi^{M}(8) } \pmb{\sigma}^{K}_{\alpha\delta}\right]^{00}\nonumber\\
&=-\left[  \pmb{\mathcal{G}}_{\delta\mu}(7,9)  \frac{ \delta \pmb{\mathcal{G}}^{-1}_{\mu\nu}(9,10)  }{\delta \Phi^{M}(8) } \pmb{\mathcal{G}}_{\nu\alpha}(10,7^{+}) \pmb{\sigma}^{K}_{\alpha\delta} \right]^{00}\nonumber\\
&=\left[  
\pmb{\mathcal{G}}_{\delta\mu}(7,9)  
\mathbf{\Lambda}^{M}_{\mu\nu}(9,10;8)
\pmb{\mathcal{G}}_{\nu\alpha}(10,7^{+})
\pmb{\sigma}^{K}_{\alpha\delta}
\right]^{00}.
\end{align}
Where we used the fact that ${G}_{\delta\alpha}\equiv{\mathcal{G}}^{00}_{\delta\alpha}$ to capture both the ordinary and the anomalous contributions to the polarizability that arise from the matrix products of Nambu Green’s functions. Now we may express the screened interaction as
\begin{widetext}
\begin{align}
w^{LJ}(6,1)=
&\varepsilon^{-1}_{LI}(6,3)v^{IJ}(3,1)\nonumber\\
=&\left(  \delta(6,3)\delta_{LI} +
 v^{KL}(7,6)\frac{\delta\braket{\mathcal{T}n^{K}(7)}}{\delta \pi^{I}(3)}
\right)v^{IJ}(3,1)\nonumber\\
=& v^{LJ}(6,1)   +
 v^{KL}(7,6)\left(  \frac{\delta\braket{\mathcal{T}n^{K}_{e}(7)}}{\delta \pi^{I}(3)} + \frac{\delta\braket{\mathcal{T}n^{K}_{n}(7)}}{\delta \pi^{I}(3)}
\right)v^{IJ}(3,1)\nonumber\\
=& v^{LJ}(6,1)   +
 v^{KL}(7,6) \frac{\delta\braket{\mathcal{T}n^{K}_{e}(7)}}{\delta \pi^{I}(3)} v^{IJ}(3,1) + v^{KL}(7,6) \frac{\delta\braket{\mathcal{T}n^{K}_{n}(7)}}{\delta \pi^{I}(3)} v^{IJ}(3,1)\nonumber\\
=& v^{LJ}(6,1)
+
v^{KL}(7,6) 
\frac{\delta\braket{\mathcal{T}n^{K}_{e}(7)}}{\delta \Phi^{M}(8)}  
\frac{\delta \Phi^{M}(8)}{\delta \pi^{I}(3)}
v^{IJ}(3,1) + v^{KL}(7,6) \frac{\delta\braket{\mathcal{T}n^{K}_{n}(7)}}{\delta \pi^{I}(3)} v^{IJ}(3,1)\nonumber\\
=& v^{LJ}(6,1)
+
v^{LK}(6,7) 
p^{KM}_{e}(7,8)  
\varepsilon^{-1}_{MI}(8,3)
v^{IJ}(3,1)+ v^{LK}(6,7) \frac{\delta\braket{\mathcal{T}n^{K}_{n}(7)}}{\delta \pi^{I}(3)} v^{IJ}(3,1)\nonumber\\
=& v^{LJ}(6,1)
+
v^{LK}(6,7) 
p^{KM}_{e}(7,8)  
w^{MJ}(8,1) 
+ v^{LK}(6,7) \frac{\delta\braket{\mathcal{T}n^{K}_{n}(7)}}{\delta \pi^{I}(3)} v^{IJ}(3,1).
\end{align}
\end{widetext}
We can further simplify this expression by solving for $w$, to obtain:
\begin{align}\label{eq:screenedeq}
w^{LJ}(8,1)& =
w^{MJ}_{e}(8,1)
+
w^{MK}_{e}(8,7) 
\frac{\delta\braket{\mathcal{T}n^{K}_{n}(7)}}{\delta \pi^{I}(3)}
v^{IJ}(3,1).
\end{align}
To evaluate the response of the nuclei to the external perturbing field, we recognize that it is equivalent to response of the total density to a perturbation in the external source field that couples to the nuclei. That is,
\begin{align}\label{eq:nucliresponse}
\frac{\delta\braket{\mathcal{T}n^{K}_{n}(7)}}{\delta \pi^{I}(3)}&=-\braket{\mathcal{T} \Delta n^{K}_{n}(7) \Delta  n^{I}(3) }=\frac{\delta\braket{\mathcal{T} n^{I}(3)}}{\delta J^{K}(7)}
\end{align}
where we defined $\Delta n^{I}_{n}(3) = n^{I}_{n}(3) - \braket{ n^{I}_{n}(3) }$ and used $\hat{J}=\int d3 J^{I}(3) n^{I}_{n}(3)$. We then can expand in terms of the electronic degrees of freedom, 
\begin{align}\label{eq:nuclifluc}
&\frac{\delta\braket{\mathcal{T} n^{I}(3)}}{\delta J^{K}(7)}=\frac{\delta\braket{\mathcal{T} n^{I}_{e}(3)}}{\delta J^{K}(7)}+\frac{\delta\braket{\mathcal{T} n^{I}_{n}(3)}}{\delta J^{K}(7)}\nonumber\\
&=
\frac{\delta\braket{\mathcal{T} n^{I}_{e}(3)}}{\delta \Phi^{N}(9)}
\frac{\delta \Phi^{N}(9) }{\delta \braket{\mathcal{T}n^{A}(10)}}
\frac{\delta \braket{\mathcal{T}n^{A}(10)}}{\delta J^{K}(7)}
+
\frac{\delta\braket{\mathcal{T} n^{I}_{n}(3)}}{\delta J^{K}(7)}\nonumber\\
&=
p^{IN}_{e}(3,9) 
v^{NA}(9,10)
\frac{\delta \braket{\mathcal{T}n^{A}(10)}}{\delta J^{K}(7)}
+
D^{IK}(3,7)\nonumber\\
&=\varepsilon^{-1}_{e~IA}(3,10) D^{AK}(10,7)
\end{align}
where we defined the nuclei fluctuation response as
\begin{align}
D^{AK}(10,7) = -\braket{\mathcal{T} \Delta n^{A}_{n}(10) \Delta  n^{K}_{n}(7) }.
\end{align}
Finally, we combine Eq.~\ref{eq:screenedeq}, Eq.~\ref{eq:nucliresponse}, and Eq.~\ref{eq:nuclifluc} to obtain:
\begin{align}
&w^{MJ}(8,1) =
w^{MJ}_{e}(8,1)
+
w^{MK}_{ph}(8,1)
\end{align}
where 
\begin{align}
&w^{MJ}_{e}(8,1) =\varepsilon^{-1}_{e~ML}(8,6)v^{LJ}(6,1)
\end{align}
and
\begin{align}
w^{MJ}_{ph}(8,1) &=
w^{MK}_{e}(8,7) 
\varepsilon^{-1}_{e~IA}(3,4) D^{AK}(4,7)
v^{IJ}(3,1),\nonumber\\
&=w^{MK}_{e}(8,7) 
D^{KA}(7,4)
v^{JI}(1,3)
\varepsilon^{-1}_{e~IA}(3,4),\nonumber \\
&=w^{MK}_{e}(8,7) 
D^{KA}(7,4)
\varepsilon^{-1}_{e~JI}(1,3) 
v^{IA}(3,4),\nonumber\\
&=w^{MK}_{e}(8,7) 
D^{KA}(7,4)
w^{JA}_{e}(1,4),
\end{align}
thus yielding Eq.~\ref{eq:HedinScreenedInt}.
In going from the second to the third line, we used the equality
\begin{align}
v(1-p_{e}v)^{-1}=(1-vp_{e})^{-1}v,
\end{align}
known from standard Green's function manipulations\cite{gonis1992green} to obtain the final expression in agreement with Refs.~\onlinecite{hedin1970effects,giustino2017electron}. 
We have thus derived the set of self-consistent Gor'kov-Hedin equations. 

\section{The Phonon Propagator}\label{appendix:Phonon}
The derivation of the phonon propagator closely follows Baym\cite{baym1961field} and Giustino\cite{giustino2017electron} except for the inclusion of spin dependent interactions. Since the commutator of the nuclear momentum $\hat{P}_{\kappa p}^{j}$ and nuclear displacements $\hat{\Delta}\tau_{\kappa p}^{i}$ is
\[
\left[ \hat{\Delta}\tau_{\kappa p}^{i}, \hat{P}_{\kappa^\prime p^\prime}^{j} \right] = i\hbar \delta_{ij}\delta_{\kappa\kappa^\prime}\delta_{pp^\prime},
\]
with 
\[
\left[ \hat{P}_{\kappa p}^{i}, \hat{P}_{\kappa^\prime p^\prime}^{j} \right] = 0 =
\left[ \hat{\Delta}\tau_{\kappa p}^{i}, \hat{\Delta}\tau_{\kappa^\prime p^\prime}^{j} \right],
\]
we must go to the second order Heisenberg equation of motion for the nuclear displacements 
\begin{align}
\frac{d^2 \hat{\Delta}\tau_{\kappa p}^{i}(\tau_1)}{d\tau_{1}^2}=\left[\mathcal{H},\left[\mathcal{H}, \hat{\Delta}\tau_{\kappa p}^{i}(\tau_1) \right] \right],
\end{align}
mirroring Newton's second Law for a mass-spring system. To proceed, we expand the nuclear density in small nuclear displacements:
\begin{align}
n^{I}_{n}(\mathbf{r}\tau) &= \sum_{\kappa p} X^I_{\kappa}\delta(\mathbf{r}-\pmb{\tau}^{0}_{\kappa p}-\pmb{\hat{\Delta}\tau}_{\kappa p}(\tau))\nonumber\\
&\approx\sum_{\kappa p} X^I_{\kappa}
\left[ 
\delta(\mathbf{r}-\pmb{\tau}^{0}_{\kappa p}) 
-
\pmb{\hat{\Delta}\tau}_{\kappa p}(\tau)\cdot\mathbf{\nabla}_{\mathbf{r}}\delta(\mathbf{r}-\pmb{\tau}^{0}_{\kappa p})\right.\nonumber\\
&\left.+
\frac{1}{2}\pmb{\hat{\Delta}\tau}_{\kappa p}(\tau)\cdot\mathbf{\nabla}_{\mathbf{r}}\left(\mathbf{\nabla}_{\mathbf{r}}\delta(\mathbf{r}-\pmb{\tau}^{0}_{\kappa p})\cdot\pmb{\hat{\Delta}\tau}_{\kappa p}(\tau)\right)
\right.\nonumber\\
&\left.-\dots\right]
\end{align}
where we used the definition of $n^{I}_n$ and taken $\pmb{\tau}_{\kappa p}(\tau)=\pmb{\tau}^0_{\kappa p}+\mathbf{\Delta}\tau_{\kappa p}(\tau)$, where $\pmb{\tau}^0_{\kappa p}$
is the average position of nucleus $\kappa$ in unit cell $p$ $(\equiv\braket{\mathcal{T}\{ \mathbf{\tau}_{\kappa p}(t_3)\}})$ and $\mathbf{\Delta}\tau_{\kappa p}(t_3)$ are the nuclear displacements about $\pmb{\tau}^0_{\kappa p}$. After computing the commutator we obtain:
\begin{widetext}
\begin{align}
M_{\kappa}\frac{d^2 \hat{\Delta}\tau_{\kappa p}^{i}(\tau_1)}{d\tau_{1}^2}=X^{I}_{\kappa}\left[ -\nabla^{i}_{\mathbf{r}}\delta(\mathbf{r}-\pmb{\tau}^{0}_{\kappa p})
+\nabla^{i}_{\mathbf{r}}\left(   \mathbf{\nabla}_{\mathbf{r}}\delta(\mathbf{r}-\pmb{\tau}^{0}_{\kappa p})\cdot \hat{\mathbf{\Delta}}\tau_{\kappa p}(\tau_1)  \right)
\right]v^{IJ}(\mathbf{r},\mathbf{r}^\prime)n^{J}_{(\kappa p)}(\mathbf{r}^\prime\tau_1)
\end{align}
\end{widetext}
where $n^{J}_{(\kappa p)}(\mathbf{r}^\prime\tau_1)=n^{J}_{e}(\mathbf{r}^\prime\tau_1)+n^{(\kappa p)J}_{n}(\mathbf{r}^\prime\tau_1)$ and $n^{(\kappa p)J}_{n}(\mathbf{r}^\prime\tau_1)=\sum_{\kappa^{\prime\prime}p^{\prime\prime}\neq\kappa p}X^{J}_{\kappa^{\prime\prime}p^{\prime\prime}}\delta(\mathbf{r}^\prime-\pmb{\tau}^{0}_{\kappa^{\prime\prime}p^{\prime\prime}})\delta(\tau,\tau_1)$ is the total density and equilibrium nuclei density excluding nuclei $\kappa p$, respectively. Multiplying by $\hat{\Delta}\tau_{\kappa^\prime p^\prime}^{j}(\tau_2)$ from the right and taking the time ordered ensemble average, we may write the resulting equation of motion in terms of the phonon Green's function defined as:
\begin{align}
\mathfrak{D}^{ij}_{\kappa p ,\kappa^\prime p^\prime}(\tau_1,\tau_2)=-\braket{  \mathcal{T} \left\{ \hat{\Delta}\tau_{\kappa p}^{i}(\tau_1) \hat{\Delta}\tau_{\kappa^\prime p^\prime}^{j}(\tau_2) \right\}  }.
\end{align}
With these definitions, we obtain the equation of motion of the phonon Green’s function
\begin{widetext}
\begin{align}\label{eq:phononEQM}
&M_{\kappa}\frac{d^2}{d\tau_{1}^2}\mathfrak{D}^{ij}_{\kappa p ,\kappa^\prime p^\prime}(\tau_1,\tau_2)=\delta_{\kappa\kappa^\prime}\delta_{ij}\delta(\tau_1,\tau_2)\nonumber\\
&-X^{I}_{\kappa p}
\left[
-\nabla^{i}_{\mathbf{r}}\delta(\mathbf{r}-\pmb{\tau}^{0}_{\kappa p})v(\mathbf{r},\mathbf{r}^\prime)\frac{\delta \braket{  \mathcal{T} \left\{ n^{J}_{(\kappa p)}(\mathbf{r}^\prime\tau_1) \right\}  }}{\delta F^{j}_{\kappa^\prime p^\prime}(\tau_2)}
-
\nabla^{i}_{\mathbf{r}}\nabla^{l}_{\mathbf{r}}\delta(\mathbf{r}-\pmb{\tau}^{0}_{\kappa p})  \mathfrak{D}^{lj}_{\kappa p ,\kappa^\prime p^\prime}(\tau_1,\tau_2)
\braket{n^{J}_{(\kappa p)}(\mathbf{r}^\prime\tau_1)}v^{IJ}(\mathbf{r},\mathbf{r}^\prime)\right],
\end{align}
\end{widetext}
where we used the fact that 
\begin{align}\label{eq:funcdivF}
\braket{n^{J}_{(\kappa p)}\hat{\Delta}\tau_{\kappa^\prime p^\prime}^{j}(\tau_2)}=\frac{\delta \braket{  \mathcal{T} \left\{ n^{J}_{(\kappa p)}(\mathbf{r}^\prime\tau_1) \right\}  }}{\delta F^{j}_{\kappa^\prime p^\prime}(\tau_2)},
\end{align}
with $\hat{F}=\sum_i \int d3 F^{i}_{\kappa p}(\tau_1) \hat{\Delta}\tau_{\kappa p}^{i}(\tau_1) $. To evaluate the response of the total density to a external
perturbing force, we recognize that
\begin{align}
n^{J}_{(\kappa p)}(\mathbf{r}^\prime\tau_1)&=n^{J}_{e}(\mathbf{r}^\prime \tau_1)+n^{J}_{n}(\mathbf{r}^\prime \tau_1)\nonumber\\
&-X^{J}_{\kappa p}\delta(\mathbf{r}^\prime - \pmb{\tau}^{0}_{\kappa p}-\mathbf{\Delta}\tau_{\kappa p}(\tau_1))\nonumber\\
&\approx n^{J}(\mathbf{r}^\prime \tau_1)-X^{J}_{\kappa p}
\left[ \delta(\mathbf{r}^\prime - \pmb{\tau}^{0}_{\kappa p})\right.\nonumber\\
&\left.-\mathbf{\Delta}\tau_{\kappa p}(\tau_1)\cdot\nabla_{\mathbf{r}}\delta(\mathbf{r}^\prime - \pmb{\tau}^{0}_{\kappa p})
\right],
\end{align}
and we then can expand in terms of the electronic degrees of freedom:
\begin{widetext}
\begin{align}\label{eq:divF}
\frac{\delta \braket{  \mathcal{T} \left\{ n^{J}_{(\kappa p)}(\mathbf{r}^\prime\tau_1) \right\}  }}{\delta F^{j}_{\kappa^\prime p^\prime}(\tau_2)}
&=
\frac{\delta \braket{  \mathcal{T} \left\{ n^{J}(\mathbf{r}^\prime\tau_1) \right\}  }}{\delta F^{j}_{\kappa^\prime p^\prime}(\tau_2)}
-\cancelto{0}{
\frac{\delta \braket{  \mathcal{T} \left\{ X^{J}_{\kappa p}\delta(\mathbf{r}^\prime - \pmb{\tau}^{0}_{\kappa p}) \right\}  }}{\delta F^{j}_{\kappa^\prime p^\prime}(\tau_2)}
}
+
\frac{\delta \braket{  \mathcal{T} \left\{ \mathbf{\Delta}\tau_{\kappa p}(\tau_1) \right\}  }}{\delta F^{j}_{\kappa^\prime p^\prime}(\tau_2)}\cdot\nabla_{\mathbf{r}}\delta(\mathbf{r}^\prime - \mathbf{\tau}^{0}_{\kappa p})X^{J}_{\kappa p}\nonumber\\
&=\varepsilon^{-1JK}_{e}(\mathbf{r}^\prime\tau_1,\mathbf{r}^{\prime\prime}\tau^{\prime\prime})
\frac{\delta \braket{  \mathcal{T} \left\{ n^{K}_{n}(\mathbf{r}^{\prime\prime}\tau^{\prime\prime}) \right\}  }}{\delta F^{j}_{\kappa^\prime p^\prime}(\tau_2)}
+
\mathfrak{D}^{lj}_{\kappa p ,\kappa^\prime p^\prime}(\tau_1,\tau_2)\nabla^{l}_{\mathbf{r}}\delta(\mathbf{r}^\prime - \pmb{\tau}^{0}_{\kappa p})X^{J}_{\kappa p}
\end{align}
\end{widetext}
Inserting Eq.~\ref{eq:divF} into Eq.~\ref{eq:phononEQM} and performing many manipulations\cite{giustino2017electron}, we arrive to Eq.~\ref{eq:baym} and \ref{eq:SEphonon}.

\end{appendix}

\bibliographystyle{naturemag}
\bibliography{SC_Hedin_Refs}

\end{document}